\documentclass[aps, pra, a4paper, showpacs, twocolumn, english, 10pt]{revtex4-2}
\usepackage[T1]{fontenc}
\usepackage{babel}
\usepackage{bbm, amsthm, bm, textcomp, nicefrac, geometry, ragged2e}
\usepackage[dvipsnames]{xcolor}
\usepackage{float}
\usepackage[bbgreekl]{mathbbol}
\usepackage{graphicx, epstopdf, color, verbatim, enumitem}
\usepackage[normalem]{ulem}
\usepackage{pbox, array}
\usepackage{mathrsfs}
\usepackage{amssymb}
\usepackage{mathtools}
\usepackage{siunitx}
\usepackage{stackrel}
\usepackage[dvipsnames]{xcolor}
\usepackage[thinlines]{easytable}
\usepackage{amsmath}
\makeatletter

\usepackage[caption=false]{subfig}
\addto\captionsenglish{}

\usepackage{hyperref}
\hypersetup{
    colorlinks = true,
    linkcolor = CadetBlue,
    citecolor =  CadetBlue,
    filecolor = CadetBlue,      
    urlcolor = CadetBlue,
}

\newcommand{\ket}[1]{\left|#1\right\rangle}

\newcommand{\bra}[1]{\left\langle #1\right|}

\definecolor{brickred}{rgb}{0.8, 0.0, 0.0}

\begin{document}

\title{Localized Entanglement Purification}

\author{Katerina Stloukalova}
\author{Jorge Miguel-Ramiro}
\author{Wolfgang D\"ur}
\author{ Julius Wallnöfer}
\affiliation{Universit\"at Innsbruck, Institut f\"ur Theoretische Physik, Technikerstra{\ss}e 21a, 6020 Innsbruck, Austria}

\date{\today}

\begin{abstract}

Entanglement purification protocols are fundamental primitives in quantum communication, enabling the distillation of high-quality entanglement using only local operations and classical communication. For large multipartite states, however, existing purification schemes typically require substantial resources and become progressively inefficient as system size increases. We introduce a new type of multipartite entanglement purification, \textit{Localized Entanglement Purification} (LEP), a family of protocols that purify entanglement at the level of network regions rather than globally. By exploiting spatial noise asymmetries, LEP reduces resource consumption and enables scalable purification strategies for larger quantum systems.
\end{abstract}

\maketitle

\section{Introduction}
Quantum technologies have advanced rapidly over the past two decades, driven by improved experimental capabilities and a growing range of possibilities. A central goal is the development of quantum networks \cite{Kimble_2008,DBLP,QNT,PhysRevA.80.022339,Fang_2023}, which enable applications such as distributed quantum information processing \cite{jones2012quantum,cariolaro2015quantum,gisin2007quantum,kremer1995quantum}, distributed quantum computing \cite{Bugalho2023,8910635,cirac1999distributed,beals2013efficient,Cacciapuoti2020}, distributed sensing \cite{PhysRevLett.112.150802,PhysRevResearch.2.023052,zang2024quantum}, entanglement-based cryptographic and multiparty protocols \cite{PhysRevLett.67.661,yin2020entanglement,PhysRevA.59.1829,PhysRevA.61.042311,Li2024,BENNETT20147,RevModPhys.92.025002,Xie_2021}. All of these applications fundamentally rely on entanglement, which degrades under noise during generation, transmission and manipulation phases \cite{PhysRevLett.81.5932,PhysRevLett.88.047902}.

Several approaches exist to combat noise, including improved entanglement generation \cite{Scheel_2001,Hannegan_2021}, quantum error correction \cite{PhysRevA.55.900,Dr2007,PhysRevLett.104.180503}, and entanglement purification protocols (EPPs) \cite{PhysRevA.80.042308,PhysRevLett.110.260503}.  Quantum repeaters mitigate this degradation by using some of these elements \cite{RevModPhys.95.045006,PhysRevA.59.169,PhysRevA.85.062326}. In particular, entanglement purification \cite{Pan2001} distills high-fidelity entangled states from multiple noisier copies using only local operations and classical communication, with well-studied recurrence-based and hashing-type protocols \cite{PhysRevLett.76.722,PhysRevLett.77.2818,PhysRevA.71.062325,PhysRevLett.125.060405,PhysRevLett.127.040502}. Graph states \cite{Gottesman1997,Hein2004, hein2006entanglementgraphstatesapplications} form an important class of multipartite entangled states, with two-colorable \cite{Aschauer_2005, asch_original} subclasses including GHZ states \cite{de_Bone_2020,rozgonyi2025,Krastanov_2021}, and cluster states \cite{PhysRevLett.86.910,Krastanov_2021,Kruszynska_2006} serve as key resources for multiple communication and computation tasks, such as measurement-based quantum computation \cite{jmbqc, Briegel2009}.

Considerable effort has been devoted to improving, extending, and optimizing EPPs \cite{Fujii2009,Rozp2018,Krastanov_2019,MR2023,Salek2025,Balmaseda25,MiguelRamiro2025,Popp2026}. However, these approaches typically address symmetric noise scenarios. Here, we focus instead on naturally occurring noise asymmetries. In this work, we present the \textit{Localized Entanglement Purification (LEP)} protocol, a multipartite EPP for arbitrary graph states that specifically targets noise locally. Using small noisy auxiliary states, mainly GHZ states, one can partially purify the main graph state. Specifically, one purifies the noise associated with particular qubits, making LEP particularly suitable for mitigating asymmetric noise. This contrasts with the standard method for purifying two-colorable graph states \cite{asch_original}, which uses full-size copies of the graph state and serves as the baseline for analyzing the performance of our method. In particular, the success probability of a purification step for the TCP decreases exponentially with the system size, which our approach avoids. 

We analyze various ways to employ the elementary localized entanglement purification protocol in a number of scenarios for different achievable fidelities and the resource cost (in terms of channel uses to generate the initial states). Our key findings include:

\begin{itemize}
    \item We show that the Localized Entanglement Purification (LEP) protocol allows for the purification of local noise in any graph state.
    \item We observe an advantageous behavior in terms of reduced resource consumption when targeting strongly asymmetric noise, which persists even in the presence of additional symmetric noise.
    \item We demonstrate the adaptability of the LEP protocol by combining it with other entanglement purification protocols to achieve higher fidelities in noisier regimes.
\end{itemize}

This work is structured as follows. The relevant background on graph states is simulated in Sec.~\ref{sec:graph states}, followed by the error model description in Sec.~\ref{sec:err} to express noisy graph states. Next, we establish the methodology for two-colorable purification protocols in Sec.~\ref{sec:back}, and for the localized entanglement purification protocol in Sec.~\ref{sec:LEP}. The effects of these two methods on noisy graph states are simulated in Sec.~\ref{sec:methods}, where we analyze and compare them under different circumstances. Initially, we investigate a strongly asymmetric noise scenario applied to a linear cluster state in Sec.~\ref{sec:only_1_z}. We then consider more complex settings by incorporating gate noise and imperfect initial states in Sec.~\ref{sec:3n}. Furthermore, we investigate in more detail the parameter regime in which our approach is advantageous by defining a \textit{Fixed Target Fidelity} (Sec.~\ref{TF}) or by attempting to achieve the highest fidelity with a given number of \textit{Fixed Total Resources} (Sec.~\ref{sec:TR}). We also demonstrate the applicability of our approach to a wider class of graph states by considering more involved combinations of EPPs applied to a two-dimensional cluster state in Sec.~\ref{sec:2d_cluster} before concluding and providing an outlook in Sec.~\ref{sec:outlook}.

\section{Background}
\label{sec:back}
In this section, we establish the theoretical foundation and essential terminology used throughout the paper.

\subsection{Graph states}
\label{sec:graph states}
This section provides an overview of graph states Refs. \cite{Hein2004, hein2006entanglementgraphstatesapplications}. A graph, $G = (V, E)$, is a mathematical structure consisting of a set of vertices $V = \left\{ 1,2,..., N \right\}$ and a set of edges $E \subseteq V \times V$, where each edge represents a connection between a pair of vertices. In the context of graph states,  vertices correspond to qubits, while edges correspond to bipartite interactions between qubits. 

A graph state $\ket{G}$ is defined by initializing $N$ qubits in the $\ket{+} = \frac{1}{\sqrt{2}}(\ket{0}+\ket{1})$ state, following the application of a two-qubit controlled-Z gates $CZ = \text{diag}(1,1,1,-1)$ on each pair of qubits connected by an edge
\begin{equation}
    \ket{G} = \prod_{(i,j) \in E}^{} CZ_{i,j}\ket{+}^{\otimes N}.
\end{equation}

The entanglement structure of the system can be fully characterized by an $N$-set of commuting correlation operators,
\begin{equation}
\label{eq:K}
    K_i = \sigma_{x}^{(i)} \prod_{j \in N(i)} \sigma_{z}^{(j)},
\end{equation}
where $\sigma_{x}^{(i)}$ is the Pauli-$X$ operator acting on qubit $i \in V$, and $\sigma_{z}^{(j)}$ is the Pauli-$Z$ operator acting on qubit $j$. Here, $N(i) \subset V$ denotes the neighborhood of vertex $i$, i.e., the set of all vertices $j$ such that $(i,j) \in E$. Hence, the form of each operator $K_i$ is determined by the adjacency matrix of the graph.

The corresponding graph state $\ket{G}$ is defined as the unique simultaneous eigenstate of all operators $K_i$ with eigenvalue $+1$, i.e. $ K_i \ket{G} = \ket{G}$  $ \forall i \in V$. The operators $ \left\{ K_i \right\}$, therefore, generate the stabilizer group of the graph state and uniquely specify $\ket{G}$ up to global phase.

Instead of using the graph state's vector description, we can use its stabilizers to characterize the properties. A complete graph state basis $\left\{ \ket{\boldsymbol{\mu}} \right\}$ can be constructed by applying local Pauli-$Z$ operations ($\sigma_{z}^{(l)}$) to the graph state $\ket{G}$ with the basis state labeled by a bit string $\boldsymbol{\mu} = \mu_{1},\mu_{2},...,\mu_{N}, \text{ with } \mu_l \in \left\{ 0,1\right\}$. The basis elements are given by
\begin{equation}
    \label{eq:graph state basis}
    \ket{\boldsymbol{\mu}} = \left( \prod_{l=1}^{N} \left( \sigma_{z}^{(l)} \right)^{\mu_{l}} \right) \ket{G}.
\end{equation}
From now on, the notation $\ket{\boldsymbol{\mu}}$ should always be understood as a graph state basis state. Each basis state is an eigenstate of the corresponding correlation operators
\begin{equation}
    \label{eq:Koper}
    K_{i}\ket{\boldsymbol{\mu}} = (-1)^{\mu_{i}} \ket{\boldsymbol{\mu}},
\end{equation}
where the bit string  $\mu_{i}$ labels the eigenvalues $\pm 1$ of the stabilizer generators $K_i$. The direct relations between $\mu_i$ and the eigenvalues of the correlation operators imply that each graph state basis state can also be uniquely identified by considering only the eigenvalues.

\subsection{Noise models}
\label{sec:err}
In any realistic setting, systems are exposed to undesirable interactions with the environment, which can be modeled as noise maps acting on the quantum system.
Thus, we introduce the noise models used to evaluate our purification protocols.

\subsubsection{Noise channels}
A frequently considered type of noise is local white (or fully depolarizing) noise $\mathcal{E}_{w}$, which is described acting on qubit $i$ with an error parameter $p_w$ by
\begin{equation}
\label{pw}
    \mathcal{E}^{(i)}_{w}(p_w) \rho 
    = p_w\,\rho 
      + \frac{1-p_w}{4}\sum_{i=0}^{3} 
      \sigma^{(i)}_{a}\,\rho\,\sigma^{(i)}_{a},
\end{equation}
where $\sigma_{a}^{(i)} \in \{\mathbb{1}, X, Y, Z\}$ are the Pauli operators.

Similarly, a local Pauli-Z noise model (dephasing), which contributes to the loss of coherence in a quantum system, is given by
\begin{equation}
\label{pz}
     \mathcal{E}^{(i)}_{z}(p_{z}^{(i)}) \rho 
     = p_{z}^{(i)}\,\rho 
       + (1-p_{z}^{(i)})\,\sigma^{(i)}_{z}\,\rho\,\sigma^{(i)}_{z},
\end{equation}
with an error parameter $p_{z}^{(i)}$.

\subsubsection{Noise sources}

We consider a setting in which a multipartite graph state is generated and distributed among spatially separated parties. This distribution requires transmitting qubits through quantum channels, which introduces noise affecting each qubit individually.

One of the main sources of noise acting on a multipartite graph state, therefore, arises from the distribution process, i.e., the transmission of qubits through imperfect quantum channels or other systematic errors in their handling. We model this by applying uniformly distributed local white noise $\mathcal{E}^{(i)}_{w}$ to each qubit, as defined in Eq.~\eqref{pw}.

Another source of noise originates from the storage of qubits in quantum memories of varying quality or for different durations. We model this by local Pauli-$Z$ noise $\mathcal{E}^{(i)}{z}$, given in Eq.~\eqref{pz}, which is applied selectively to individual qubits with qubit-dependent strength $p{z}^{(i)}$.

In addition to the uniformly distributed local white noise baseline described above, we incorporate local Pauli-$Z$ noise into the system, resulting in the following initial state model:

\begin{equation}
\label{init}
\rho_{init} = \prod_{i \in I \subset V}^{} \mathcal{E}_{z}^{(i)}(p_{z}^{(i)}) \prod_{i \in  V}^{} \mathcal{E}_{w}^{(i)}(p_{w}^{(i)}) \ket{G}\!\bra{G}
\end{equation}

Next, we model noisy operations such as CNOT gates (which are used frequently by the EPPs) by local white noise acting on the control and target qubits $(c,t)$ followed by the perfect CNOT operation $U^{c\to t}_{CNOT}$. Therefore, we describe the map of a noisy CNOT gate as 
\begin{equation}
     \mathcal{M}^{c\to t}_{CNOT}(p_g)\rho 
     = U^{c\to t}_{CNOT}
       \left(
       \mathcal{E}^{(t)}_{w}(p_g)\,
       \mathcal{E}^{(c)}_{w}(p_g)\,\rho
       \right)
       U^{\dagger c\to t}_{CNOT},
\end{equation}
with \textit{gate noise} parameter $p_g$. The gate noise is repeatedly applied during the use of the purification protocol, whereas the initial noise map is only acts once at the beginning.

\subsection{Multipartite entanglement purification for two-colorable graph states}
\label{sec:tcp}
Two-colorable entanglement purification (TCP) is a multipartite recurrence purification protocol that considers a scenario in which $N$ distinct parties share many identical copies of a noisy multipartite entangled state associated with a two-colorable graph \footnote{ Two-colorable graph is a graph, whose vertices are colored only using two colors, such that no two neighboring vertices are colored the same.}, as described in \cite{Aschauer_2005,asch_original}.

The TCP protocol proceeds in successive purification rounds, each acting on two identical copies using local operations and classical communication (LOCC)\footnote{Although the protocol involves two-qubit gates such as CNOTs, as shown in Fig. \ref{fig:TCP}, these are performed locally within each party’s laboratory. The protocol, therefore, falls within the LOCC framework.}. At each purification step, two identical copies undergo graph-dependent multilateral CNOT (MCNOT) operations that correlate their error information without directly revealing it. Subsequently, one of the two copies is locally measured and discarded, with the measurement outcomes being classically communicated and checked against graph-specific parity conditions. If these conditions are satisfied, the remaining copy is retained, thereby increasing fidelity probabilistically. To obtain a high-fidelity multipartite graph state from initially noisy resources, this distillation procedure must be repeated several times.



\subsubsection{TCP protocol details}
\label{ssec:tcpfid}

A graph state is two-colorable if the vertices of the associated Graph $G = (V,E)$, can be partitioned in two disjoint color sets $V_A \text{ and } V_B$, with  $V = V_A \cup V_B, V_A \cap  V_B = \emptyset$, such that no vertices from qubits from the same color-set can't be connected $ \forall a \in V_A: N(i) \subseteq V_B \text{ and } \forall b \in V_B: N(b) \subseteq V_A$ \cite{Aschauer_2005,asch_original}.

We define the graph state basis by splitting the bit string $\boldsymbol{\mu}$ into two parts corresponding to the two color settings $\ket{\boldsymbol{\mu}} = \ket{\boldsymbol{\mu}_A,\boldsymbol{\mu}_B}$
which contains the local basis contributions from each color set. Meaning that the stabilizer eigenvalues $\boldsymbol{\mu}_A,\boldsymbol{\mu}_B$, associated with color sets $A$ and $B$, respectively, encode the presence of local error information on each qubit \cite{Aschauer_2005,asch_original}.

Mathematically, the protocol steps can be summarized as follows. Consider a two-colorable mixed state that is diagonal in the graph state basis
\begin{equation}
\label{eq:diagonaltcp}
    \rho_{\boldsymbol{\mu}} = \sum_{\boldsymbol{\mu}_{A}, \boldsymbol{\mu}_{B}}^{} \lambda_{\boldsymbol{\mu}_{A}, \boldsymbol{\mu}_{B}} \ket{\boldsymbol{\mu}_{A}, \boldsymbol{\mu}_{B}} \bra{\boldsymbol{\mu}_{A}, \boldsymbol{\mu}_{B}},
\end{equation}
where the $\lambda_{\boldsymbol{\mu}_{A}, \boldsymbol{\mu}_{B}}$ coefficients represent classical probabilities associated with each basis element in the mixed state. Since the noise is Pauli-diagonal, the graph state maps into a classical mixture over the graph state basis states. 

In this representation, the basis element $\boldsymbol{\mu}_{A},\boldsymbol{\mu}_{B} = \boldsymbol{0},\boldsymbol{0}$ corresponds to the ideal graph state. Hence, the first diagonal element $\lambda_{0,0}$ equals the overlap of the mixed state with the desired pure state $\ket{\Phi}$, resulting in
\begin{equation}
    F = \bra{\Phi}\rho_{\boldsymbol{\mu}}\ket{\Phi} = \lambda_{\boldsymbol{0},\boldsymbol{0}},
\end{equation}
where $F$ is the state fidelity \cite{Nielsen_Chuang_2024} with respect to the ideal target state $\ket{\Phi}$. 

We denote the second or auxiliary copy as $\rho_{\boldsymbol{\nu}}$ (with the graph state basis states denoted as $\ket{\boldsymbol{\nu}} = \ket{\boldsymbol{\nu}_A,\boldsymbol{\nu}_B}$), which is initially described in the same way, resulting in a tensor product of these two noisy copies $\rho_{0} = \rho_\mu \otimes \ \rho_\nu$. 

To the input state, we apply the two-qubit multilateral CNOT operation (MCNOT). The designation of the source and target qubits, which are used for the MCNOT, is given by the TCP purification sub-protocols ($P_1$ and $P_2$), each acting on a color class of the graph state and differing in the direction of the MCNOT application relative to the graph state's coloring.

\begin{figure} 
    \centering
    \includegraphics[width=0.48\textwidth]{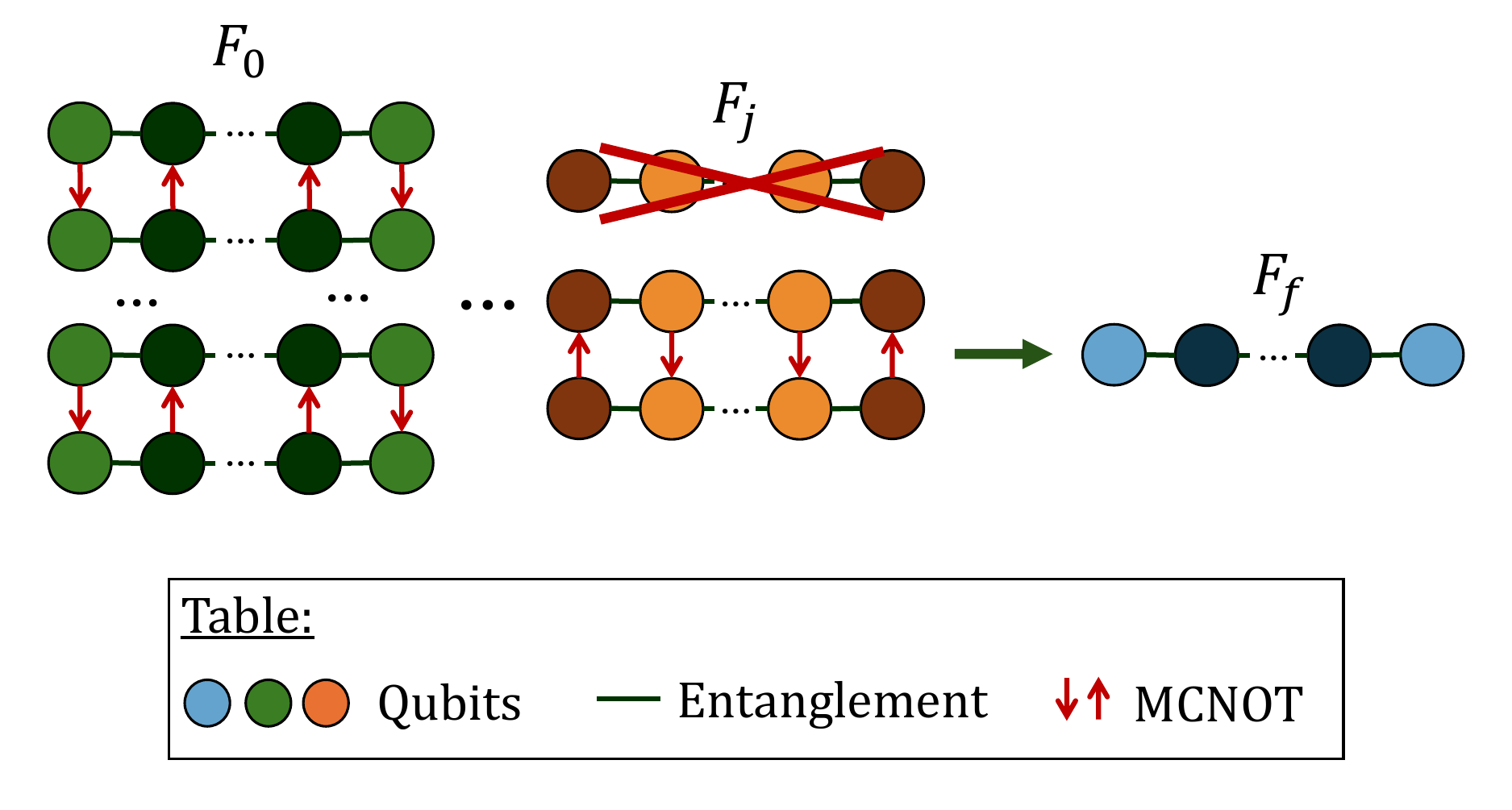}
    \caption{ \textbf{Illustration of the TCP protocol.} Multiple noisy copies of an $N$-linear graph state with an initial fidelity $F_0$. The arrows denote the multilateral (MCNOT) directions associated with the sub-protocols $P_1$ and $P_2$, while the color shading distinguishes the two color sets. After measurement and post-selection, copies failing the purification step are discarded, whereas successful copies with increased fidelity $F_j$ are retained and used in subsequent rounds. After several purification rounds, a single copy with the highest achievable fidelity $F_f$ remains.}
    \label{fig:TCP}
\end{figure}

\textit{Sub-protocol $P_1$.--} The first sub-protocol, $P_1$ employs the following gate-direction pattern. First, CNOTs are applied to $\ket{\boldsymbol{\mu}_{A}, \boldsymbol{\mu}_{B}},\ket{\boldsymbol{\nu}_{A} \boldsymbol{\nu}_{B}}$, where the qubits in color set $A$ use $\rho_{\boldsymbol{\nu}}$ as the control and $\rho_{\boldsymbol{\mu}}$ as the target. The set $B$ applies the gate in the opposite direction, which is described as
\begin{equation}
\label{eq:p1mcnot}
    U_{MCNOT}^{P_1} = \prod_{i\in V_A}^{} U_{CNOT}^{\boldsymbol{\nu}_{i}\to \boldsymbol{\mu}_{i}}\prod_{l\in V_B}^{}U_{CNOT}^{\boldsymbol{\mu}_{l}\to \boldsymbol{\nu}_{l}},
\end{equation}
and is also illustrated in Fig.~\ref{fig:TCP}.
One can use the CNOT–Pauli commutation rules, the stabilizers in Eq. \eqref{eq:K}, and the action of the MCNOT on the basis state anti-commutes relations in Eq. \eqref{eq:Koper} and rewrite the state as
\begin{equation}
\label{eq:GSB}
    \ket{\boldsymbol{\mu}_{A}, \boldsymbol{\mu}_{B}} \ket{\boldsymbol{\nu}_{A}, \boldsymbol{\nu}_{B}}\xrightarrow{} \ket{\boldsymbol{\mu}_{A}, \boldsymbol{\mu}_{B} \oplus \boldsymbol{\nu}_{B}} \ket{\boldsymbol{\nu}_{A} \oplus \boldsymbol{\mu}_{A}, \boldsymbol{\nu}_{B}},
\end{equation}
where $\oplus$ denotes bit-wise addition modulo 2. Highlighting the fact that the MCNOTs maps the tensor product of the graph state basis to another product of the graph state basis.



We now perform a measurement determined by the stabilizer operators in Eq.~\eqref{eq:K} for each sub-color class. Specifically, for sub-protocol $P_1$, qubits belonging to the color set $V_A$ are measured in the $X$-basis, while qubits in the color set $V_B$ are measured in the $Z$-basis. The corresponding stabilizer relations define the parity conditions used for post-selection.

If all the measurement outcomes correspond to the $+1$ stabilizer eigenvalue,i.e., it satisfies $\boldsymbol{\nu}_{A} \oplus \boldsymbol{\mu}_{A} = 0$, the purification is declared as successful. If the condition is not fulfilled, the remaining copy is discarded, and the purification step is deemed unsuccessful, as shown in Fig. \ref{fig:TCP}.

After a successful round of purification, the coefficients of the new mixed state are updated accordingly
\begin{equation}
\label{eq new denmat}
    \widetilde{\rho}_{\boldsymbol{\mu}} = \sum_{\boldsymbol{\widetilde{\mu}}_A,\boldsymbol{\widetilde{\mu}}_B}^{} \widetilde{\lambda}_{{\boldsymbol{\widetilde{\mu}}_A,\boldsymbol{\widetilde{\mu}}_B}} \ket{{\boldsymbol{\widetilde{\mu}}_A,\boldsymbol{\widetilde{\mu}}_B}}\bra{\boldsymbol{\widetilde{\mu}}_A,\boldsymbol{\widetilde{\mu}}_B},
\end{equation}
where $\boldsymbol{\widetilde{\mu}}_A = {\boldsymbol{\mu}}_A$ and $\boldsymbol{\widetilde{\mu}}_B = {\boldsymbol{\mu}}_B \oplus {\boldsymbol{\nu}}_B$, such that the summation effectively runs over all possible bit strings ${\boldsymbol{\nu}}_B$. This is visible in the coefficient normalization
\begin{equation}   
\label{eq:coeff new}
\widetilde{\lambda}_{\widetilde{\boldsymbol{\mu}}_{A},\widetilde{\boldsymbol{\mu}}_{B}} = \frac{1}{2K}\sum_{(\boldsymbol{\mu}_{B},\boldsymbol{\nu}_{B})|\boldsymbol{\mu}_{B}\oplus\boldsymbol{\nu}_{B} = \widetilde{\boldsymbol{\mu}}_{B}}^{} \lambda_{\boldsymbol{\mu}_{A}\boldsymbol{\mu}_{B}}\lambda_{\boldsymbol{\nu}_{A}\boldsymbol{\nu}_{B}}.
\end{equation}
The $2K$ factor serves both as a normalization constant and as the protocol's success probability, denoted by $p_{j}$, where $j$ indexes the chosen sub-protocol.

The overall effect of $P_1$ sub-protocol, if successful, is to amplify all the coefficients $\lambda_{0, \boldsymbol{\mu}_B}$.

\textit{Sub-protocol $P_2$.--} The complementary sub-protocol $P_2$ is analogously defined to purify with respect to the second color, which is done following similar steps, with the difference that the MCNOT operations are applied in the reversed direction, as shown in Eq. \eqref{eq:p1mcnot}, and the measurement bases are switched as well. As a result, for qubits in set $A$, the controls are in $\rho_{\boldsymbol{\mu}}$ and target qubits are in $\rho_{\boldsymbol{\nu}}$, while for set $B$ the roles are switched. The corresponds to relabeling $\boldsymbol{\mu}_A \to \boldsymbol{\nu}_A \text{ and } \boldsymbol{\nu}_B \to \boldsymbol{\mu}_B$. 

Eq. \eqref{eq:GSB} is modified such that the graph state basis labels become $\boldsymbol{\widetilde{\mu}}_B = {\boldsymbol{\mu}}_B$ and $\boldsymbol{\widetilde{\mu}}_A = {\boldsymbol{\mu}}_A \oplus {\boldsymbol{\nu}}_A$, which in turn determines the corresponding update of the coefficients in the new density matrix, as given in Eq. \eqref{eq new denmat}—the sub-protocol $P_2$ compensates for the noise redistribution induced by the $P_1$ operation. As such, it makes sense to apply $P_1$ and $P_2$ in an alternating fashion as a recurrence protocol, i.e., starting from many copies, the output states of one purification step are used as the input for the next.

\subsubsection{TCP resource evaluation}
We define channel uses as a metric to evaluate the resource requirements of the EPP. Distributing an $N$-qubit graph state among distinct parties requires either sending $N-1$ qubits from a central source through quantum channels or distributing Bell pairs over channels and connecting them locally via gate operations.


The TCP protocol requires an identical copy per purification round; the counting is straightforward. For the case of \(\mathcal{K}\) successful rounds, the average resource consumption is
\begin{equation}
\label{eq:restcp}
    R_{\mathrm{TCP}} = L \prod_{k=1}^{\mathcal{K}} \frac{2}{p_{j_{k}}}, \qquad j \in \{1,2\}.
\end{equation}
Here \(L\) denotes the number of quantum channels used to generate the main graph state (e.g., \(L = N-1\) for a linear cluster state of size \(N\)). The average number of resources is given by \(2/p_{j_{k}}\), with $p_{j_k}$ denoting the success probability for a specific sub-protocol \(P_{j_k}\), which is applied in the purification round $k$. Moreover, the success probability $p_{j}$ for a given protocol $P_{j}$ generally decreases with increasing system size, as a large number of stabilizer conditions must be simultaneously satisfied. Note that the required resources scale exponentially with the number of qubits $N$ \cite{hein2006entanglementgraphstatesapplications,asch_original}. This protocol serves as the baseline for comparison with the other EPPs we analyze in this work.


\section{Localized Entanglement purification protocol}
\label{sec:LEP}

We introduce here our proposed purification framework, denoted as \textit{Localized Entanglement Purification (LEP)}, which can be applied to arbitrary graph states. Unlike conventional approaches (such as the TCP protocol reviewed above) that treat all qubits uniformly \cite{Aschauer_2005,Kruszynska_2006}, and whose success probability drops exponentially, our method can selectively target individual qubits, making it particularly effective against asymmetric noise. By locally mitigating noise on specific qubits, the framework significantly increases success probability and reduces the number of required copies. 

The LEP protocol can be viewed as a pumping-like purification that targets a noisy main target state, iteratively purifying it using freshly prepared smaller auxiliary states (GHZ states). The target copy is kept while repeatedly consuming additional auxiliary states to transfer error information and probabilistically increase its fidelity. Successful rounds gradually improve the quality of the target state with moderate resource overhead.

\subsection{Operational setting}
Consider an arbitrary graph state defined on a vertex set $V$ including all qubits. This set is partitioned into three disjoint subsets, $V = V_T \dot\cup V_{N_T} \dot\cup V_{T_{0}}$. This partition depends on the selection of a target qubit $T$, which forms the subset $V_T$. In this paper, we restrict ourselves to a single target qubit.

For the set $V_{N_T}$, we select the neighborhood of the target qubit $T$, which is given by $N(T) = \{ b \in V \mid \{T,b\} \in E \}$ and consists of all qubits directly connected to $T$.  The remaining set $V_{T_{0}}$ comprises all qubits, which are not connected to $T$. We emphasize that the partitioning is adaptive and depends on the choice of the target qubit $T$. 

In contrast to purification protocols that require identical copies of the graph state, LEP employs smaller auxiliary states tailored to the selected target qubit $T$. Specifically, LEP protocol uses a GHZ state comprising only the target qubit $T$ and its neighborhood $N(T)$, thus leaving the set $V_{T_{0}}$ empty. 

\subsubsection{LEP protocol details}
Consider an initial mixed state of the main graph state, whose diagonal form in the graph state basis is given by
\small
\begin{equation}
\label{eq:LEPp}
    \rho_{\boldsymbol{\mu}} = \sum_{{\mu}_{T}, \boldsymbol{\mu}_{T_{N}}, \boldsymbol{\mu}_{T_{0}} }^{} \lambda_{{\mu}_{T}, \boldsymbol{\mu}_{T_{N}},\boldsymbol{\mu}_{T_{0}}} \ket{{\mu}_{T}, \boldsymbol{\mu}_{T_{N}}, \boldsymbol{\mu}_{T_{0}}} \bra{{\mu}_{T}, \boldsymbol{\mu}_{T_{N}},\boldsymbol{\mu}_{T_{0}}},
\end{equation}
\normalsize
where the bit strings ${\mu}_{T}$, $\boldsymbol{\mu}_{T_N}$, and $\boldsymbol{\mu}_{T_0}$ label the stabilizer eigenvalues associated with the target qubit, its neighborhood, and the remaining qubits of the graph, respectively. The coefficients $\lambda_{{\mu}_{T}, \boldsymbol{\mu}_{T_{N}}, \boldsymbol{\mu}_{T_{0}}}$ denote the corresponding classical probability distribution. Since the basis element $\mu_{T}, \boldsymbol{\mu}_{T_N}, \boldsymbol{\mu}_{T_0} = 0,\boldsymbol{0},\boldsymbol{0}$ corresponds to the ideal graph state, this provides us with the fidelity defined as $F=\lambda_{0,0,0}$. This is exactly the same as the fidelity described for the TCP protocol in Sec. \ref{ssec:tcpfid}.

The auxiliary state is described by a mixed state $\rho_{\boldsymbol{\nu}}$ that has a form similar to the two-colorable graph state basis given in Eq. \eqref{eq:diagonaltcp}, with the restriction that the stabilizer eigenvalues are defined only on the reduced set of qubits ($V_T, V_{N_T}$) comprising the auxiliary state. In particular, the qubit set associated with the $T$ qubit corresponds to ${\nu}_{T}$, while the neighborhood group is labeled by $\boldsymbol{\nu}_{T_N}$, resulting in $\ket{\boldsymbol{\nu}_{T}, \boldsymbol{\nu}_{T_{N}}}$. This is very convenient since the LEP protocol aims to manipulate and correlate the stabilizer eigenvalues of the $T$ qubit and its neighborhood while leaving the remaining qubits unaffected. 

We defined the diagonal graph state basis for the main graph state and for a single auxiliary copy. Next, we apply the multilateral CNOT (MCNOT) operations to purify the main graph state $\rho_{\boldsymbol{\mu}}$ using the reduced auxiliary copy $\rho_{\boldsymbol{\nu}}$. These operations follow the mapping $\rho_{\boldsymbol{\mu}} \to \rho_{\boldsymbol{\nu}}$ for the set $V_{T}$ and the reversed mapping $\rho_{\boldsymbol{\nu}} \to \rho_{\boldsymbol{\mu}}$ for the set $V_{T_N}$.
The combined action of these gates on a product of graph state basis elements is given by
\begin{equation}
\label{eg:gsb pk}
\begin{gathered}
\begin{aligned}
\ket{{\mu}_{T}, \boldsymbol{\mu}_{T_{N}}, \boldsymbol{\mu}_{T_{0}}}
\ket{{\nu}_{T}, \boldsymbol{\nu}_{T_{N}}}
\end{aligned}
\\[0.5em]
\downarrow {}
\\[0.5em]
\begin{aligned}
\ket{{\mu}_{T}, \boldsymbol{\mu}_{T_{N}} \oplus \boldsymbol{\nu}_{B}, \boldsymbol{\mu}_{T_{0}}}
\ket{{\nu}_{T} \oplus \mu_{T}, \boldsymbol{\nu}_{T_{N}}}.
\end{aligned}
\end{gathered}
\end{equation}

The MCNOT operations update the neighborhood stabilizer eigenvalues of the main copy $\boldsymbol{\mu}_{T}$ and the target eigenvalue of the auxiliary copy $\boldsymbol{\nu}_{T}$, while leaving all remaining qubits in the subset $V_{T_0}$ unchanged. A more detailed derivation is provided in Appendix \ref{ap:cal}. Although these qubits do not participate directly in the purification step, their corresponding coefficients contribute to the overall normalization and may be indirectly affected by subsequent noise processes.

Following this step, measurements are performed on the auxiliary state according to the stabilizer correlation operators defined in Eq. \eqref{eq:K}. Specifically, the target qubit in the set $V_T$ is measured in the $X$-basis, while the qubits in its neighborhood $V_{N_T}$are measured in the $Z$-basis. These measurement outcomes define the parity condition used for post-selection. In the LEP framework, only a single target qubit is considered, resulting in a single post-selection condition that must satisfy ${\nu}_{T} = {\mu}_{T}$. If this condition is satisfied, the purification step is successful and the main graph state copy is retained; otherwise, all copies are discarded.

Upon post-selection on the desired measurement outcomes, the coefficient update for the new density matrix takes the form:
\begin{equation}
\widetilde{\lambda}_{{\mu}_{T},\,\widetilde{\boldsymbol{\mu}}_{T_N},\,\boldsymbol{\mu}_{T_0}}
=
\frac{1}{2K'}
\sum_{\substack{
\boldsymbol{\mu}_{T_N},\,\boldsymbol{\nu}_{T_N} \\
\boldsymbol{\mu}_{T_N} \oplus \boldsymbol{\nu}_{T_N}
= \widetilde{\boldsymbol{\mu}}_{T_N}
}}
\lambda_{{\mu}_{T},\,\boldsymbol{\mu}_{T_N},\,\boldsymbol{\mu}_{T_0}}
\,
\lambda_{{\mu}_{T},\,\boldsymbol{\nu}_{T_N}},
\end{equation}
where $2K'$ is the normalization constant as well as the protocol success probability, which differs from the success probability mentioned in Eq. \eqref{eq:coeff new}. After renormalization, the fidelity gets enhanced since post-selection favors configurations in which both copies share the same error pattern, thereby increasing the relative weight of the error-free component. Effectively, by amplifying coefficients with $\mu_T = 0$, Z-noise on qubit $T$ is suppressed.

\begin{figure}
    \centering
    \includegraphics[width=0.48\textwidth]{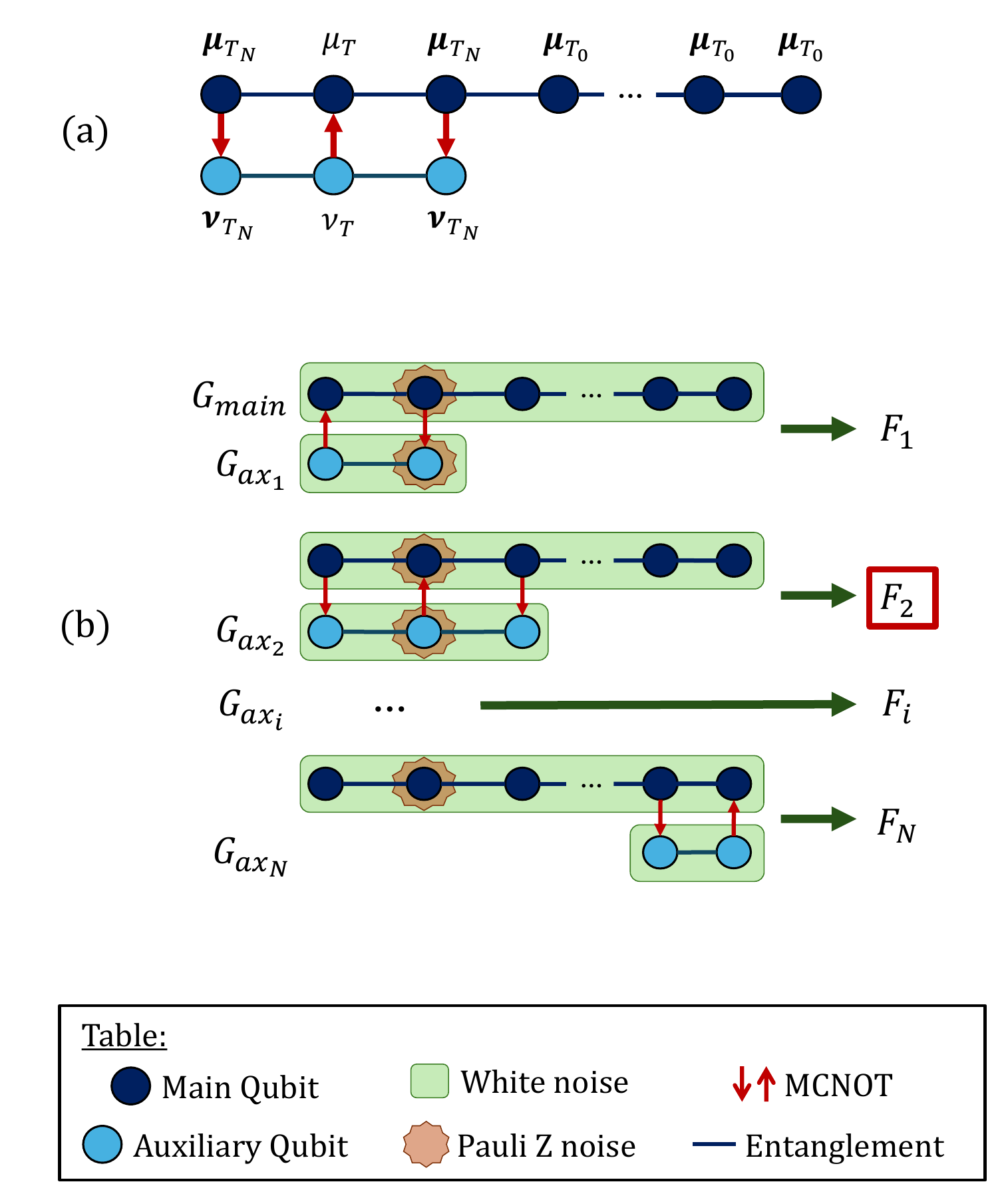}
    \caption{ \textbf{Illustration of the Localized Purification (LEP) protocol.} Example for $N$-qubit linear cluster state, with panel (a) demonstrating the LEP approach, where a target qubit $T$ is selected along with its neighboring qubits set ${T_N}$, while all other qubits in set ${T_0}$ remain inactive during MCNOT operation. The direction of the operation is determined by the target qubit $\mu_T$, and the auxiliary state is constructed from the corresponding qubit $\nu_{T}$ and the neighboring qubits $\boldsymbol{\nu}_{T_N}$. Varying position of the target qubit $T$ enables the identification of the optimal LEP strategy, as shown in (b). Starting from an initial fidelity $F_0$, all candidate auxiliary operations are virtually evaluated, and the one that maximizes the resulting fidelity (e.g., $F_2$) is selected and applied. This procedure is then iteratively repeated.}
    \label{fig:LEP}
\end{figure}

An illustrative example of an LEP framework is shown in Fig.~\ref{fig:LEP}(a). We consider an $N$ linear cluster, where each qubit is assigned to a specific subset: the target qubit $\mu_T$, its neighboring qubits $\boldsymbol{\mu}_{T_N}$, and the remaining qubits $\boldsymbol{\mu}_{T_0}$. The auxiliary state is constructed according to the location of the target qubit and, being a GHZ state, is defined on the corresponding subsets $\nu_T$ and $\boldsymbol{\nu}_{T_N}$. 

Since we defined a protocol for each choice of $T$, one can vary the target qubit $T$ to enable an adaptive purification approach. This means that one can simulate the effect of performing one purification step for each possible choice and select the one that yields the largest increase in fidelity. Note, however, that this exploration is conducted only virtually; only the variant that we select from the results is actually performed. Fig.~\ref{fig:LEP}(b) illustrated this for an $N$-qubit linear cluster state, which is subjected to white noise $\mathcal{E}_w^{\text{all}}$, and asymmetric Pauli-$Z$ noise $\mathcal{E}_{z}^{(2)}$ applied to the second qubit. It illustrates one purification round using a sequence of tailored auxiliary states ($G_{ax_1},...,G_{ax_N}$) which are subjected to the same noise as the main graph state $G_{main}$.

The procedure begins by selecting a target qubit $T$ and applying the corresponding auxiliary state $G_{ax}$, yielding an updated fidelity. This process is then repeated for all possible choices of $T$, where each auxiliary operation is virtually evaluated starting from the same initial state. The auxiliary state that produces the highest resulting fidelity is selected, providing the optimal choice for an actual experimental implementation.

In this example, the asymmetric Pauli-$Z$ noise acting on the second qubit makes $G_{ax_{2}}$ the most effective choice for the first purification round. However, the optimal choice in subsequent rounds depends on the evolving noise distribution and cannot be determined a priori 

For this reason, the protocol operates as a localized and adaptive optimization procedure, dynamically navigating these uncertainties to improve the fidelity of the target state. Optimized LEP schemes are discussed in the next section. 

\subsubsection{LEP protocol resource evaluation}
As before, we define channel uses as the main performance metric. We define one resource as a single noisy copy of a multipartite graph state. Such a state can be distributed either by transmitting $N-1$ qubits from a central source or by distributing Bell pairs and connecting them locally. 

As LEP protocol operates as a pumping-type purification scheme, the resource accounting must be adjusted to include not only the main graph state but also the variable-size auxiliary graph state used in each purification round. The average resource consumption after \(\mathcal{K}\) successful purification rounds is then given by
\begin{equation}
\label{eq:resLEP}
R_0 = L, \qquad
R_{LEP}^{n} = \frac{R_{n-1} + M_n}{p_{m_n}},
\quad n = 1,2,\dots,\mathcal{K}.
\end{equation}
Here, $R_0$ is the resource cost of preparing the main graph state. The quantity $M_n$ represents the number of quantum channels (edges) required to generate the auxiliary graph state in the $n$-th round, with a total of $\mathcal{K}$ rounds. In each round, every auxiliary state is assigned a distinct success probability $p_{m_n}$, determined by the chosen LEP strategy $m$ (for a purification strategy, $m$ will state the same for the entire purification process).

\subsection{Protocol optimization}
With the basic LEP framework established, we adopt an adaptive scheme in which the number of auxiliary states per purification round is determined by the achieved fidelity. We can further optimize the protocol by accounting for additional features.

\subsubsection{Pre-purification}
The \textit{Pre-purification} process applies only to the auxiliary graph state and aims to improve its quality before targeting the main graph purification. The protocol leverages TCP, which is known to be highly efficient for small graph states, such as GHZ states, consistent with the structure of the auxiliary graph state considered here. This pre-purification step reduces noise on the auxiliary states, which can subsequently be used to increase the efficiency of the LEP strategy. 

For instance, in Fig. \ref{fig:LEP}(b), one takes the auxiliary copy $G_{ax_2}$ and implements a TCP protocol $\alpha$ number of times with ($\alpha \in \mathbb{N}$), before using the $G_{ax_2}$ with the main copy $G_{main}$.

\subsubsection{S-$\alpha$ strategy}
\label{sub:s}
The \textit{S-$\alpha$ strategy} is the basic form of LEP, where "S" denotes \textit{single} and refers to the application of a single auxiliary graph state per purification round, thereby achieving the highest fidelity in a single round. The parameter $\alpha$ specifies the number of pre-purification TCP steps applied to the auxiliary state in each round.

For example, S-$2$ indicates that the pre-purification procedure applies twice the TCP to the auxiliary graph state, which is then used to purify the main graph state. This entire procedure is counted as a single purification round in our simulation model.

\subsubsection{C-$\alpha$ strategy}
\label{sub:c}
The \textit{C-$\alpha$ strategy} stands for the "C"ombination of purification steps. Rather than drawing conclusions from fidelity after a single purification round ($S-\alpha$), we evaluate different orderings and combinations of auxiliary graph states across two rounds and select the sequence that yields the highest final fidelity. This strategy is combined with the S-$\alpha$ approach, which selects either the one-round or two-round look-ahead, optimizes both simultaneously, and yields C-$\alpha$, which is counted as one purification step in our simulations.


For example, for the $N$-qubit cluster state, one needs $N$ auxiliary states. In the first purification round, all auxiliary states are tested, and the resulting fidelities are saved, without selecting the best outcome. A second purification round is then simulated using all auxiliary states again, resulting in $N^2$ ordered combinations. The auxiliary-state pair yielding the highest fidelity corresponds to a \textit{double} strategy. 

\subsubsection{LEP-TCP-$\alpha$ strategy}
\label{sub:hyb}
The final approach we consider is a hybrid protocol, the \textit{LEP-TCP-$\alpha$ strategy}, that combines pre-purification, LEP, and TCP. Specifically, we apply the S-$\alpha$ strategy for a limited number of rounds, determined by the strength and distribution of the asymmetric noise. Afterward, the TCP protocol is employed on the main graph state to further enhance the fidelity of the resulting state. This entire process is treated as a single purification step in our simulations.

Our goal is to demonstrate that carefully chosen combinations of auxiliary graph states can further enhance purification performance across different noise regimes, as elaborated in the next section. 

\section{Performance Analysis}
\label{sec:methods}
In this section, we present numerical results for various optimized LEP strategies and compare them with the standard TCP. During these simulations, we treat each purification step as a separate sub-protocol of TCP, and the full LEP strategy as one round, with the details described for each strategy in Secs.~\ref{sub:s}--\ref{sub:hyb}. We consider the same kind of noise for both the main and auxiliary states. To evaluate their efficiency, we compare their fidelity outputs per round and related resources. %

\subsection{Ideal 1-qubit asymmetric case}
\label{sec:only_1_z}
The first scenario we consider is an 8-node linear cluster state subjected to asymmetric noise. Specifically, local Pauli-$Z$ noise with strength $p_z = 0.7$ is applied to the first qubit of the cluster. No additional noise sources, such as white noise or gate noise, are included at this stage, allowing us to isolate and study the behavior of the LEP and TCP strategies under purely asymmetric noise conditions.


Before performing numerical simulations, we analyze the first purification step for both the TCP and LEP protocols analytically. We begin with TCP, the asymmetric  initial noisy state can be described as
\begin{equation}
\begin{split}
    \rho_{TCP} = &p_z^{(1)}\ket{\boldsymbol{0}_A,\boldsymbol{0}_B}\bra{\boldsymbol{0}_A,\boldsymbol{0}_B}+\\
    &(1-p_z^{(1)})\ket{(1,0,0,0),\boldsymbol{0}_B}\bra{(1,0,0,0),\boldsymbol{0}_B},
\end{split}
\end{equation}
where $p_z^{(1)}$ is the probability that no error occurs on qubit 1 within the corresponding color group, associated with the bit string $\boldsymbol{0}_{A/B} = (0,0,0,0)$. Since only a single Pauli-$Z$ error channel is considered, only one syndrome bit is affected, changing the bit string to $(1,0,0,0)$, which occurs with probability $p_z^{(1)}$.

Although the LEP protocol has a different qubit partitioning of the graph state compared to TCP, in this particular case, the description is similar, i.e.,
\begin{equation}
\begin{split}
    \rho_{LEP} = &p_z^{(1)}\ket{0_{T},\boldsymbol{0}_{T_N},\boldsymbol{0}_{T_0}}\bra{0_{T},\boldsymbol{0}_{T_N},\boldsymbol{0}_{T_0}}+\\
    &(1-p_z^{(1)})\ket{1_T,\boldsymbol{0}_{T_N},\boldsymbol{0}_{T_0}}\bra{1_T,\boldsymbol{0}_{T_N},\boldsymbol{0}_{T_0}},
\end{split}
\end{equation}
where qubit sub-group $\boldsymbol{\mu}_{T}$ contains only the target (noisy) qubit.

Since we are only concerned with Z-noise on qubit 1, we can track the error on that particular qubit based on the fact that $\boldsymbol{\mu}_{A} = \boldsymbol{\nu}_{B}$ and $\boldsymbol{\mu}_{T} = \boldsymbol{\nu}_{T_N}$ has to be satisfied for TCP and LEP respectively. The probability of satisfying such conditions for both protocols is $p_{j_1} = p_{m_1} = (p_z^{(1)})^2 + (1-p_z^{(1)})^2$. Since we discarded the cases where the post-selection condition is not satisfied, we re-normalized the state by updating the coefficients as
\begin{equation}
\tilde{\lambda}_{(0000),\boldsymbol{0}} = \frac{(p_{z}^{(1)})^2}{(p_{z}^{(1)})^2+(1-p_{z}^{(1)})^2}.
\end{equation}

\begin{figure} 
    \centering 
    \includegraphics[width=\columnwidth]{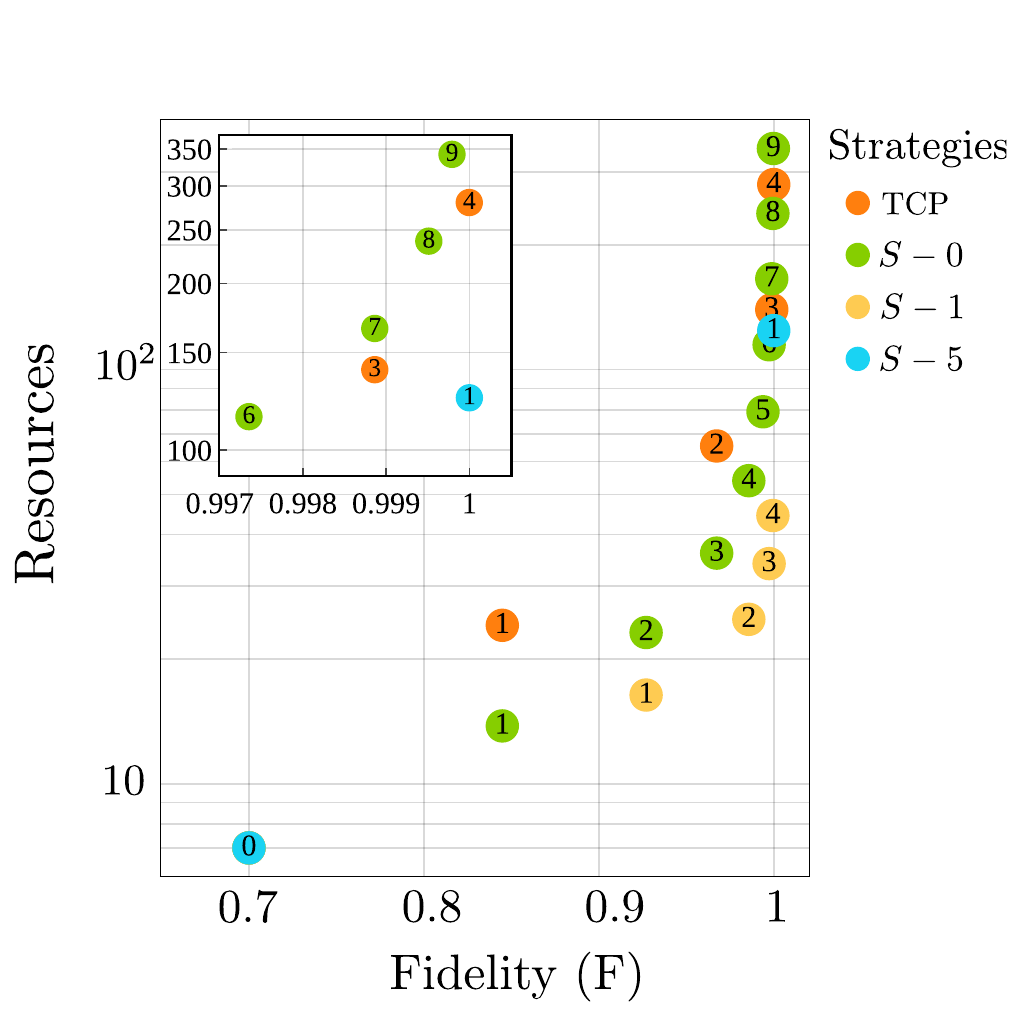}
    \caption{\textbf{Fidelity vs Resources: $\boldsymbol{Z}$-noise on qubit 1.} An 8-qubit linear cluster state subjected to Pauli-Z noise on qubit 1 (leaf) with $p_{z}^{(1)} = 0.7$, with no additional initial noise or gate noise ($p_{w} = p_{g} = 1.0$), while showcasing the effect of different purification strategies like TCP and LEP ($S-0$, $S-1$, and $S-5$).  We limit the maximum number of resources to $10^9$; any value exceeding this is declared unsuccessful.}
    \label{fig:1D-1.0-1.0}
\end{figure}

This leads to the same fidelity output value of TCP and single-step LEP ($S-0$) for the first round of purification, which is confirmed by numerical simulation in Fig. \ref{fig:1D-1.0-1.0}. However, both protocols differ in their resource consumption, which is evaluated separately for TCP and LEP in Eq.~\eqref{eq:restcp} and Eq.~\eqref{eq:resLEP}. 

In subsequent purification rounds, the two protocols no longer yield identical fidelities, reflecting the recurrence behavior of TCP and the pumping-like behavior of LEP protocol. LEP must employ a pre-purification step to compensate for the originally too-noisy auxiliary states, thereby improving the auxiliary states to a lower noise level, such as the $S-1$ or $S-5$ protocol, to maintain high fidelity.

The best outcomes in terms of fidelity and resource consumption for this case were achieved by the LEP protocol, specifically the $S-1$, approaching $F \rightarrow 1$  at its fourth call. We thus adopt $S-1$ as an optimized strategy, directly targeting the affected subsystem, yielding more efficient noise suppression and reduced resource consumption.


Eventually, the favorable behavior towards the LEP strategy can change if the 1D cluster is subjected to additional noise, as elaborated in the next section.

\subsection{Noisy 3-qubit asymmetric case}
\label{sec:3n}

Moving away from the highly idealized scenario, we consider here additional white noise and gate imperfections, and extend the asymmetric noise model to affect a larger subset of qubits.

Again, we showcase an 8-qubit linear cluster state that is now subjected to white noise  $p_w = 0.95$ and gate noise $p_g = 0.998$, while extending Pauli-$Z$ noise to qubits: 1 ($p_{z}^{(1)} = 0.81$),3 ($p_{z}^{(3)} = 0.9$), and 6 ($p_{z}^{(6)} = 0.85$).

\begin{figure}
    \centering 
    \includegraphics[width=\columnwidth]{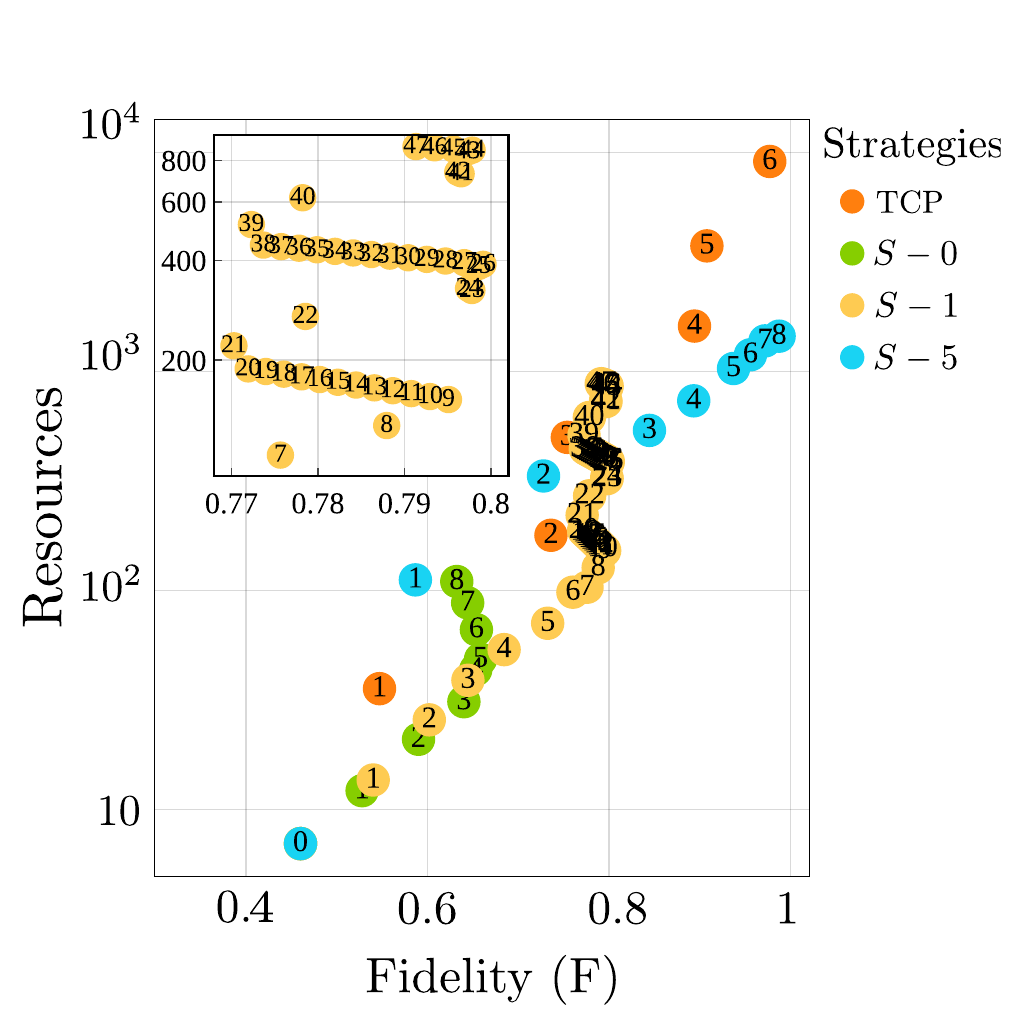}
    \caption{\textbf{Fidelity vs Resources: $\boldsymbol{Z}$-noise on qubit 1, 3, and 6.} An 8-qubit linear cluster state is subjected to Pauli-Z noise on qubits: 1 ($p_{z}^{(1)} = 0.81$),3 ($p_{z}^{(3)} = 0.9$), and 6 ($p_{z}^{(6)} = 0.85$), with initial white noise  $p_w = 0.95$ and gate noise $p_g = 0.998$, with application of different purification strategies such as: LEP (strategies $S-0$, $S-1$, and $S-5$) and TCP.}
    \label{fig:1D-0.9-0.998}
\end{figure}

Fig. \ref{fig:1D-0.9-0.998} shows numerical results for this case. The $S-0$ strategy is insufficient to fully mitigate noise and reaches its maximum fidelity earlier than in less-noisy scenarios, because its pumping scheme repeatedly relies on very noisy auxiliary states. In particular, the application of the MCNOT operation propagates errors to neighboring qubits, as described in Eq.~\eqref{eg:gsb pk}, effectively increasing the noise contribution on adjacent qubits. This redistribution of noise alters the priority order of the auxiliary graph states required in subsequent purification rounds.

To address this limitation, it is beneficial to employ pre-purification,  pushing the achievable fidelity to a higher maximum $F_{max}$, but at the cost of increased resource requirements. As illustrated in Fig. \ref{fig:1D-0.9-0.998}, the location of $F_{max}$ is not clear for the $S-1$ strategy. To clarify this behavior, we follow the auxiliary graph patterns used for purification and the noise-pattern redistribution of the main graph state. 

We observe that the $S-1$ strategy initially mitigates the noise efficiently by optimally selecting different auxiliary states. However, at the 9th call (illustrated in Fig. \ref{fig:1D-0.9-0.998}), it reaches a point where it repeatedly applies the same auxiliary graph state to the same qubit. The number of such repeated purification rounds depends on the noise strength and its location, and this targeted repetition enables the protocol to approach its maximal achievable purification performance.

The redistribution of noise becomes sufficiently symmetric that the LEP protocol can no longer mitigate it effectively. Instead, in this particular linear cluster-state scenario, the protocol converges toward the leaf (i.e., the edge of the cluster). Rather than purifying the state to increase fidelity, repeated applications of the protocol progressively degrade it, specifically the calls between 9 and 20. The noise is transferred to a neighboring qubit through the application of the MCNOT gate, introducing highly asymmetric gate-induced noise on adjacent qubits (calls 21, 24, 25). Subsequent purification steps then act on these neighboring qubits, after which the protocol converges back to the leaf (calls 22, 23, 26), forming a loop.

\begin{figure*}
    \centering
    \subfloat[]{\includegraphics[width=0.5\textwidth]
    {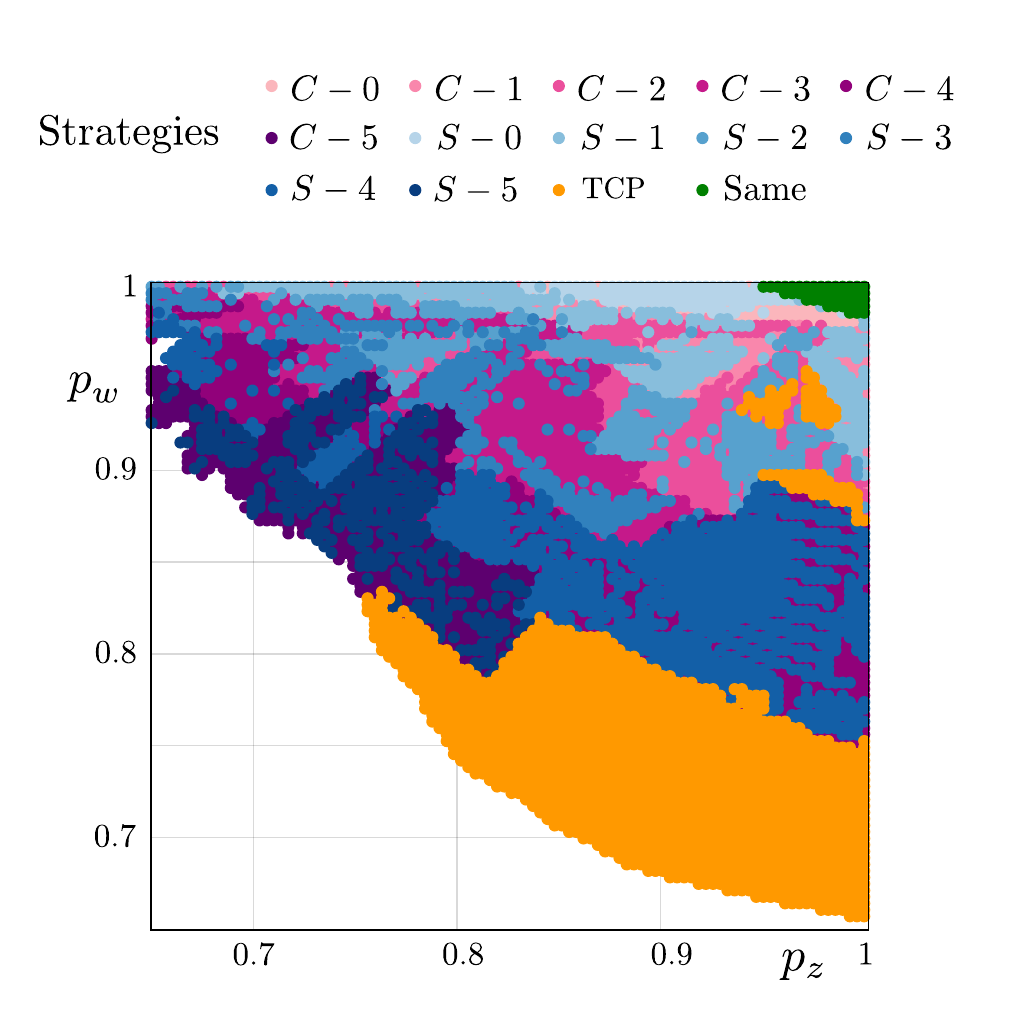}}
    \hfill
    \subfloat[]{\includegraphics[width=0.5\textwidth]{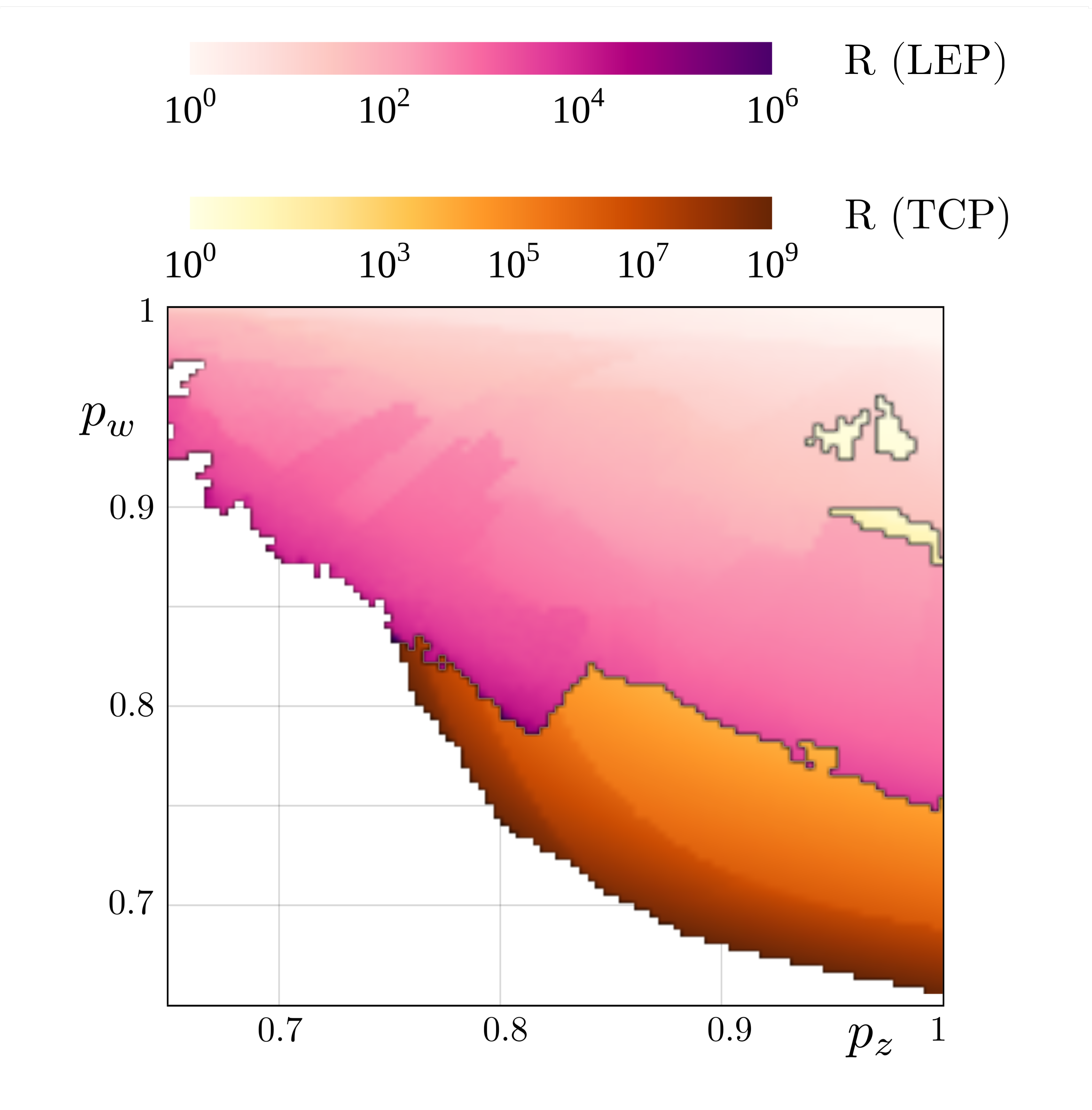}}
    \caption{\textbf{Fixed Target Fidelity ($F_T$ = 0.90) for an 8-node linear cluster state. } We consider this case to be subjected to asymmetric noise, where $Z$ noise with strength $p_z$ acts on qubits 1 and 6, while the white-noise parameter $p_w$ is varied simultaneously. The figure compares different LEP-based strategies, namely S-$\alpha$ and C-$\alpha$, and TCP. The S-$\alpha$ applies the auxiliary protocol with $\alpha$ pre-purification rounds. The combined strategy C-$\alpha$ consists of applying strategy $S$ once or twice, and selecting the option that yields the best performance relative to the TCP strategy. Sub-figures (a) and (b) show results for fixed target fidelity \textbf{$F_T$ = 0.90}. Sub-figure (a) presents outcomes of different strategies for $F_T$ = 0.90. For each noise configuration, we identify the strategy that achieves the fixed target fidelity $F_T=0.90$ with the least amount of resource consumption as the winning strategy. If the initial fidelity is already equal to or exceeds the fixed target fidelity, the result is labeled as \textit{same}. The amount of resources used for $F_T=0.90$ per TCP (orange) and LEP (pink) strategies is displayed in (b).}
    \label{fig:TF}
\end{figure*}


Due to the high resource cost, continuing until $F_{max}$ is reached may not always be desirable. Note, for instance, how in Fig. \ref{fig:1D-0.9-0.998} the relative gain fidelity $G_{\%}$ \footnote{To quantify the shift between the two highest-fidelity points, we define the relative gain $G_\%$ as the percentage increase in fidelity from the first maximum \((F_k,R_k)\) to the second maximum \((F_l,R_l)\),
\[
    G_\% = \left(  \frac{F_l - F_k}{F_k} \right)\times 100.
\]
With  $k \neq l, \text{ label distinct protocol calls with } l>k \text{ and } k,l\in K$, where $K$ is defined as the number of calls. This equation can also be adapted for resources. } between call 9 and call 26 is only 0.5 \%, whereas the corresponding increase in resource consumption between these two operation points is 155 \%. In this case, higher-order calls are marginal relative to the substantial increase in resources, yielding little or no practical advantage. 


Although one-time pre-purification $S-1$ achieves higher fidelity than $S-0$, it still does not approach a near-perfect state, as seen in TCP over 6 rounds. For that reason, we pre-purify 5 times ($S-5$), which applies less noisy auxiliary states to achieve a near-perfect state at a lower cost than TCP.


\begin{figure*}
    \centering
    \subfloat[]{\includegraphics[width=0.45\textwidth]{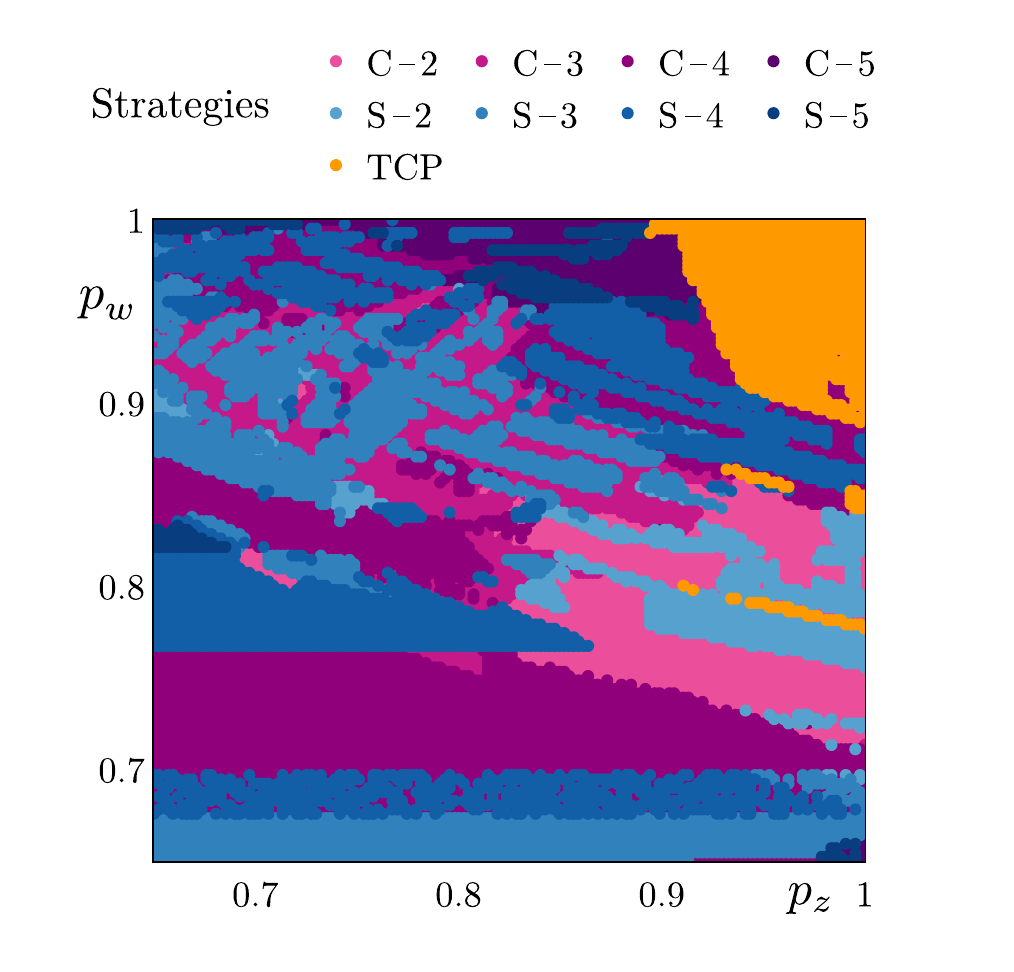}}
    \hfill
    \subfloat[]{ \includegraphics[width=0.45\textwidth]{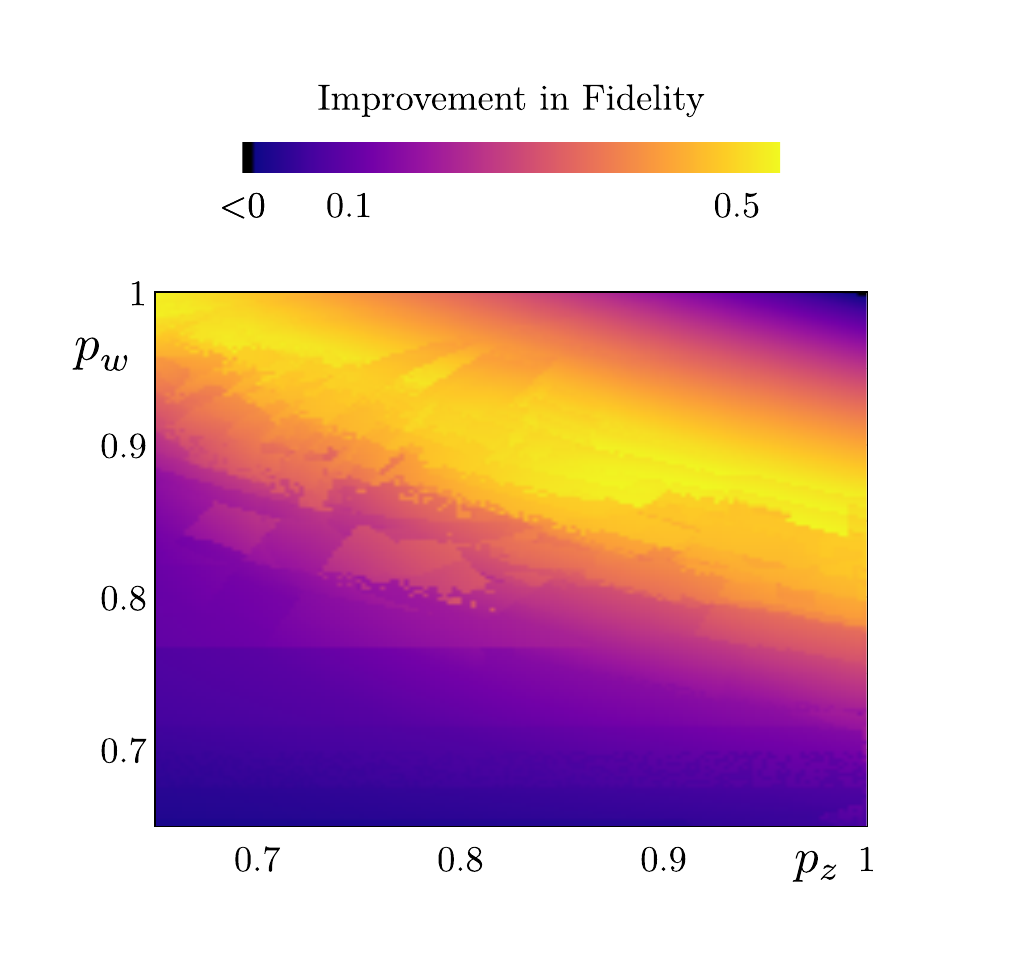}}
    \caption{\textbf{Fixed Total Resources (TR = 1000) for An 8-node linear cluster state. } We consider an asymmetric noise scenario in which $Z$ noise with strength $p_z$ acts on qubits 1 and 6, while the white-noise parameter $p_w$ is varied simultaneously. The figure compares LEP-based strategies, $S$-$\alpha$ and $C$-$\alpha$, with TCP. Strategy $S$-$\alpha$ applies the auxiliary protocol with $\alpha$ pre-purification rounds, while the combined strategy $C$-$\alpha$ selects between one or two applications of $S$ to achieve the best performance relative to TCP. Sub-figures (a) and (b) illustrate the Fixed Total Resources framework with a resource limit of \textbf{TR = 1000}. The sub-figure (a) shows the winning strategy that maximizes the achieved fidelity, while sub-figure (b) shows the corresponding fidelity gain relative to the initial state in a fixed noise regime.}
    \label{fig:TR}
\end{figure*}

\subsection{Fixed Target Fidelity}
\label{TF}
So far, we have considered scenarios with strong asymmetry, i.e., where we would expect the LEP protocol to be advantageous. To better understand the parameter regime in which this advantage holds, we fix the \textit{target fidelity (TF)}, a concept that identifies the most suitable purification strategy across a range of noise parameter values by selecting the one that minimizes resource consumption.


We illustrate this analysis for a fixed target fidelity of $F_T=0.90$, varying the white-noise parameter $p_w$ of the 8-qubit linear cluster state. An equal Pauli-$Z$ noise parameter $p_z$ is applied to qubits 1 and 6, while the gate noise is fixed at $p_g = 0.998$.

We use this noise configuration to identify the most effective strategy, i.e., the one with the lowest resource consumption, as illustrated in Fig. \ref{fig:TF}(a). The strategies considered are $S-\alpha$ and $C-\alpha$ with $\left\{ \alpha \in \mathbb{Z} | 0\le \alpha \le  5\right\}$.

To achieve such a fixed target fidelity $F_T$, we use linear interpolation between two consecutive purification steps $n$ and $n+1$, whose fidelities satisfy $F_{n} \le TF \le F_{n+1}$, where $n$ denotes the number of purification rounds. Due to the discrete nature of the purification steps, the $F_T$ may not be reached exactly. We therefore construct an effective state by probabilistically mixing the outputs of these two steps,
\begin{equation}
\label{linf}
    \rho_{n'} = p\rho_{n}+(1-p)\rho_{n+1},
\end{equation}
where $p$ is the mixing ratio. This corresponds to randomly applying either $n$ or $n+1$ purification rounds, allowing the effective fidelity to be tuned continuously. The corresponding average resource consumption is then given by 
\begin{equation}
\label{linr}
    R_{TF} = pR_{n}+(1-p)R_{n+1}.
\end{equation}

We show in Fig. \ref{fig:TF}(a) the strategies of type $C-\alpha$ under the pink color, with a gradual color shading indicating the number of pre-purification rounds applied to the auxiliary state. Similarly, strategies of type $S-\alpha$ are shown in blue, with the color intensity reflecting the corresponding value of $\alpha$. 

On the other hand, the TCP and ``same'' strategies are represented by a single color. Note that ``same'' (shown in green) indicates that no purification is required, as the initial fidelity satisfies $F_0 > TF$, corresponding to a regime with minimal or no noise applied to the initial state. By alternating between different $p_w$ and $p_z$ values, thereby affecting the initial fidelity, in a regime in which $F_0 < F_T$, requiring purification.

The LEP strategy dominates in low white noise regimes and across both weak and strong asymmetric Pauli-$Z$ noise. In contrast, at higher white-noise levels, more pre-purification rounds are required, as the use of too noisy auxiliary states limits the effectiveness of the pumping procedure and necessitates additional noise reduction before efficient purification. Eventually, a high-noise regime is reached, favoring the TCP strategy.

We observe that two distinct scenarios emerge for TCP. One corresponds to a moderately noisy state that favors TCP over LEP due to high-fidelity requirements and \textit{sweet spots}. These types of spots are expected to occur for both TCP and LEP at different target fidelities whenever the parameter regime favors a particular strategy. A second scenario occurs at high white-noise levels, where achieving the fixed target fidelity requires substantial resources. LEP accumulates excessive noise, further exacerbated by the noise distribution introduced by the MCNOT gates, rendering mitigation infeasible.

Although this scenario is idealized with respect to resources, realistic implementations are constrained by finite resources. Accordingly, we include Fig. \ref{fig:TF}(b), which illustrates the resource consumption of the optimal strategy across different noise regimes. We distinguish only two groups: orange for the TCP approach and pink for the LEP and its strategies, $S-\alpha$ and $C-\alpha$. 

The LEP protocol does not reach high-noise regimes as TCP does; such regimes require at least $10^{6}$ resources. 
Overall, the results exhibit a highly unpredictable strategy pattern, largely driven by the sensitivity to parameter variations. Additionally, the precise appearance of the pattern also depends on the chosen value $F_T$. The LEP strategy requires orders of magnitude fewer resources than TCP; this difference is highlighted in Appendix~\ref{additional app}. We also investigate the hybrid approach that combines TCP and LEP and present the results in Appendix \ref{ap:TF}.


\subsection{Fixed Total Resources}
\label{sec:TR}
The \textit{Fixed Total Resources (TR)} framework enforces resource limits and identifies which strategies can be applied across different noise regimes to achieve the highest fidelity, thereby showing the fidelity improvement achieved by the winning strategy.

The setting for defining the optimal strategy is the same as in the previous section, in which we iterate over different values of $p_w$ and $p_z$. To properly evaluate such $TR$ for any strategy and its corresponding fidelity $F_{TR}$, we employ linear interpolation as described in Sec.~\ref{TF}. First, Eq.~\eqref{linr} is used to determine the corresponding resource values, followed by Eq.~\eqref{linf} to obtain the resulting fidelities.



The resulting optimal strategies are shown in Fig. \ref{fig:TR}(a), which identifies the strategy that achieves the highest fidelity under the imposed resource constraint. The TCP strategy demonstrates the expected behavior by covering the symmetric region. However, coverage is influenced by $TR$; to expand TCP territory, we would require substantially more resources because TCP's resource requirement increases exponentially with each purification step. We can conclude that the LEP strategy also dominates in symmetric regions.

In the asymmetric noise regime, only LEP appears and produces the winning strategy with the highest fidelity. However, the behavior of the pre-purification rounds is not straightforward to predict. In some noise regimes, only a few pre-purification rounds are sufficient to achieve high fidelity, whereas in other regions we require more rounds to more effectively mitigate noise.

While Fig. \ref{fig:TR}(a) identifies the strategy that performs best in each region, Fig. \ref{fig:TR}(b) highlights where purification is actually beneficial in terms of fidelity improvement. In that case, we display a fidelity difference between the initial fidelity at given noise parameters and the final fidelity.


In the almost pure state regime, the difference between the pre- and post-purification states is negative. The state here is already close to ideal, and the purification process effectively de-purifies due to the additional gate noise introduced by the MCNOT operations, thereby reducing fidelity. As the noise level increases, purification becomes increasingly necessary and yields greater fidelity improvements. However, once the initial state becomes excessively noisy, the resulting improvement in fidelity begins to decline, indicating a limit to efficient purification, since the initial state is no longer purifiable or even entangled.

Under these conditions and within this specific framework, we include additional simulations of a hybrid strategy, which are elaborated in Appendix \ref{ap:tr}.

\subsection{2D cluster states}
\label{sec:2d_cluster}
\begin{figure}
    \centering 
    \includegraphics[width=\columnwidth]{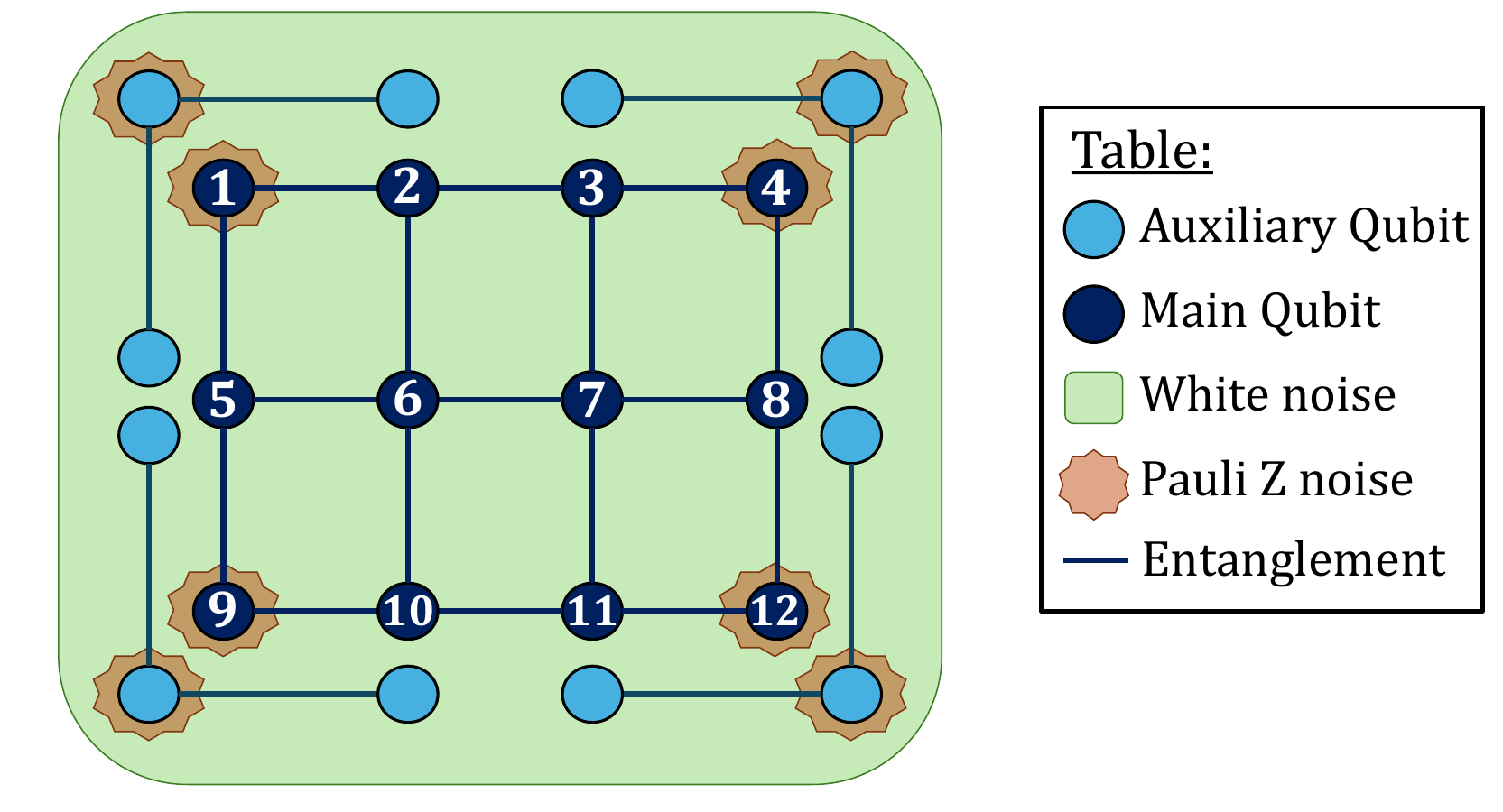}
    \caption{\textbf{Illustration of 2D cluster state.} Example of a 12-qubit two-dimensional cluster state illustrating the qubit labeling scheme. The dark blue numbered qubits represent the main graph state, which is affected by uniform local white noise, with additional Pauli-$Z$ noise applied to selected qubits. The auxiliary qubits, which form the auxiliary graph state, implement the first four purification steps that target asymmetric noise locally at the corner qubits.}
    \label{fig:2D}
\end{figure}

\begin{figure}
    \centering 
    \includegraphics[width=\columnwidth]{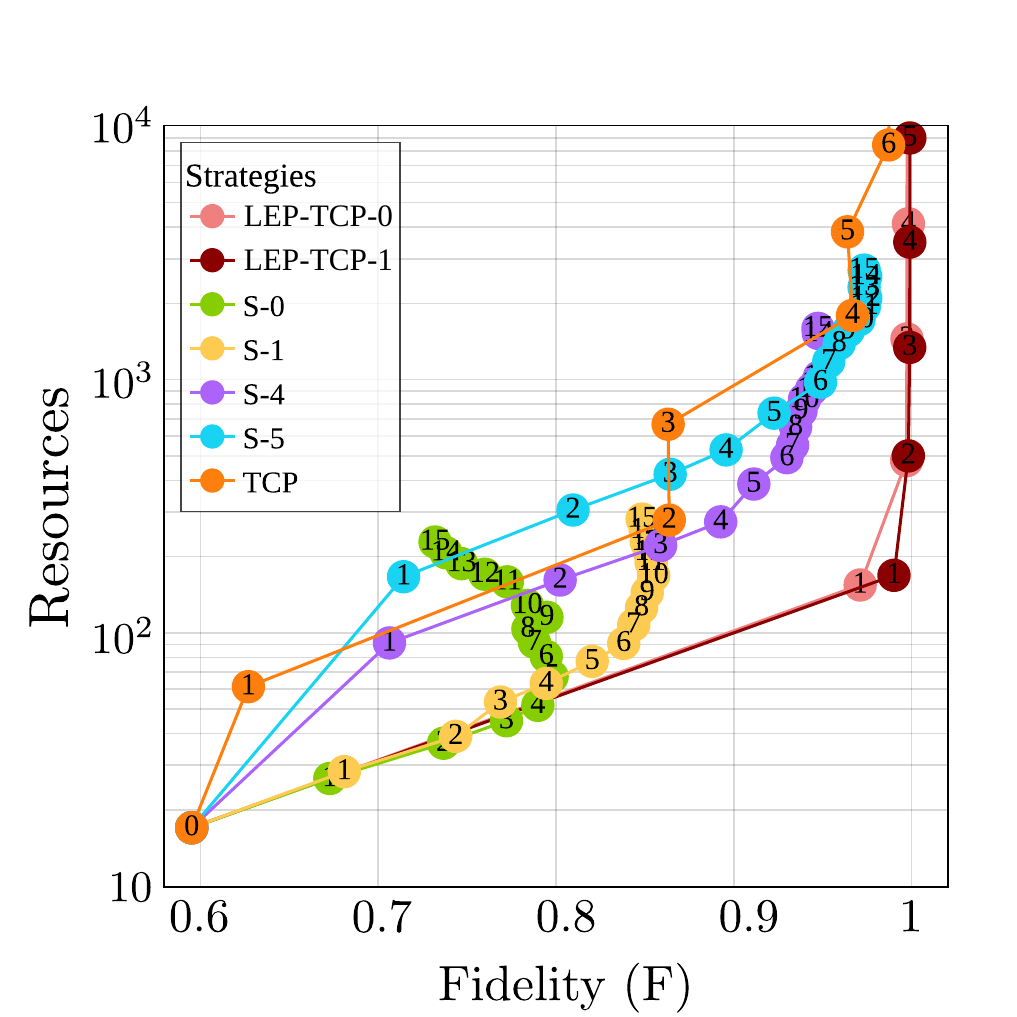}
    \caption{\textbf{Fidelity vs resources for 2D cluster state.} We consider a 12-qubit two-dimensional cluster state subjected to white noise with strength $p_w$ = 0.98, gate noise  = 0.998, together with asymmetric $Z$-noise distributed to the corners, thus affecting qubits: $p_{z}^{(1)} = 0.9$, $p_{z}^{(4)} = 0.85$, $p_{z}^{(9)} = 0.95$, and $p_{z}^{(12)} = 0.98$. The figure compares the LEP strategy in its S-$\alpha$ variant, the TCP strategy, and a hybrid LEP–TCP–$\alpha$ approach, where $\alpha$ denotes the number of pre-purification rounds.}
    \label{fig:2d4cs}
\end{figure}

To show that our method is applicable to any graph state, we extend our analysis beyond linear cluster states and introduce a simple 3x4 2D cluster state comprising 12 qubits, which is shown in Fig.~\ref{fig:2D}. We present a simple illustrative example in which the corner qubits are affected by additional Pauli-$Z$ noise on top of the local white noise acting on all qubits. In this scenario, the first four auxiliary states are primarily used to target the corner qubits due to the strong asymmetric noise. However, this behavior is not universal and depends on the relative strengths of the Pauli-$Z$ noise, gate noise, and white noise. This dependence will be examined in more detail in the example below for specific choices of noise parameters. We investigate different strategies, such as $S-\alpha$, TCP, and $LEP-TCP-\alpha$ (hybrid), and plot their corresponding fidelities and resource requirements in Fig. \ref{fig:2d4cs}. We quantify the resource cost by the number of edges in the 2D cluster state, since each such edge requires the distribution and consumption of a Bell pair.

The simplest strategy, $S-\alpha$, exhibits its typical behavior, requiring fewer resources to achieve higher fidelity in the presence of asymmetric noise. However, even with unlimited resources and additional pre-purification steps, this strategy does not approach the nearly perfect state as TCP does. This limitation arises because $S-\alpha$ effectively redistributes the noise symmetrically across the system, thereby exhausting its purification potential earlier than the other strategies and reusing very noisy auxiliary states.

Meanwhile, TCP can eventually reach a near-perfect state, but it requires substantial resources, whereas the hybrid LEP-TCP-$\alpha$ strategy avoids this. Because $S-\alpha$ initially targets asymmetric noise, thereby resulting in symmetric noise redistribution, employing TCP enhances overall fidelity and consumes the most resources in the first purification step. 

We remark that, as described in the hybrid strategy in Sec. \ref{sub:hyb}, the protocol should be applied based on the strength and number of qubits affected by asymmetric noise.

In this noise configuration, the number of LEP rounds of $S-0$ or $S-1$ used for the hybrid approach is determined by the \textit{effective} noise distribution rather than by the mere presence of asymmetric noise on individual qubits.

For this particular example, we introduce 4 qubits that are more strongly affected by asymmetric Pauli-$Z$ noise. The 12th qubit exhibits the same level of noise as the white noise and eventually does not require a dedicated local purification step. Since the MCNOT uniformly amplifies the initial noise across neighboring qubits during LEP on the specific target qubit. This effect increases the overall noise level, causing a symmetric distribution, and may necessitate purification of the neighboring qubits rather than purifying the 12th qubit locally.

Consequently, the LEP protocol is applied only 3 times, followed by a final purification step using the TCP protocol, employing either sub-protocol $P_1$ or $P_2$, and completing one full purification step for the LEP-TCP-$\alpha$ strategy. 

The effect of this hybrid approach is most pronounced in the first purification round, which is the most resource-intensive, yielding a near-perfect state in which certain strategies become inaccessible in subsequent rounds. Even when accounting for pre-purification of the auxiliary states during the LEP rounds, this approach enables a modest further increase in the achievable fidelity. Thus, the hybrid strategy is the most promising in this case.

\section{Summary and Outlook}
\label{sec:outlook}
We have introduced a multipartite entanglement purification protocol for graph states that targets noise locally using a small GHZ state as an auxiliary state. We demonstrated that this localized entanglement purification (LEP) protocol effectively targets local noise while requiring fewer resources, e.g., compared to the TCP protocol we used as a baseline, which requires a full second copy of the main graph state. 

We introduced multiple variants of this protocol and demonstrated that purification remains robust enough to be beneficial even in the presence of other sources of noise, although the exact behavior depends on the parameters.

We have shown that the LEP protocol is particularly suitable when the system noise is highly asymmetric, as this situation highlights the LEP's ability to target a specific contribution to the overall noise model. Furthermore, we have demonstrated that the resource efficiency of our approach is particularly pronounced when only part of the noise needs to be mitigated to reach a given fixed target fidelity, or when resources are limited. In addition, the LEP strategy is flexible with respect to graph-state size and dimensionality, making it applicable to large-resource states, such as 2D cluster states.

Due to the complex trade-offs and the dependence on precise parameters, a search for an ideal strategy is likely to involve combining multiple approaches, such as the hybrid approach we discuss here. While we have highlighted many variations for using and combining the EPPs in this work, further optimizations could improve performance. However, this is likely to be challenging, as there are many potential protocol combinations to consider, e.g., a pumping-type application of TCP or a variant of the LEP that varies the number of pre-purification steps per protocol round.

On a more conceptual level, our approach could be useful for generating, distributing or maintaining large multipartite resource states, e.g., entanglement-based networks \cite{Cuquet_2012,Walln_fer_2019,Meignant_2019,prielinger2025piecemakerresourceefficiententanglementdistribution} or measurement-based quantum computation \cite{Freund_2025}. In such a network, multipartite states would need to be stored in memories for significant amounts of time, and building a second copy of the whole state may be prohibitively complex, whereas generating an auxiliary GHZ state with neighboring networks to locally reduce the noise in parts of the state is more feasible. A step in that direction could be to integrate the LEP with network simulations for protocols such as the merging-based quantum repeater \cite{morruiz2025mergingbasedquantumrepeater}, which could be used as a method to obtain a distributed multipartite entangled state.

\section*{Acknowledgments}
This research was funded in whole or in part by the Austrian Science Fund (FWF) 10.55776/P36009, 10.55776/P36010, and 10.55776/COE1. For open access purposes, the author has applied a CC BY public copyright license to any author accepted manuscript version arising from this submission. Finanziert von der Europ\"aischen Union - NextGenerationEU.

\bibliographystyle{apsrev4-2}
\bibliography{references} 

\newpage

\appendix
\section{Derivation of the LEP graph state basis}
\label{ap:cal}
We provide here more details about the MCNOT transformation for the main and auxiliary graph state $\ket{{\mu}_{T}, \boldsymbol{\mu}_{T_{N}}, \boldsymbol{\mu}_{T_{0}}}\ket{{\nu}_{T}, \boldsymbol{\nu}_{T_{N}}} = \ket{\boldsymbol{\omega}}$, where we use $\ket{\boldsymbol{\omega}}$ for a shorter description.

Firstly, we define the direction of MCNOT operations as
\begin{equation}
\label{eg:gsappendix}
U_{M} = U_{CNOT}^{\nu_{T}\to \mu_{T}} U_{CNOT}^{\boldsymbol{\mu}_{T_N}\to \boldsymbol{\nu}_{T_N}}
\end{equation}
with
\begin{equation}
\label{eg:gsb pk appendixx}
U_{CNOT} := \prod_{i \in N(T)}^{} U_{CNOT}^{\mu^{(i)}\to \nu^{(i)}}.
\end{equation}
Within the stabilizer formalism, the effect of this MCNOT operation can be evaluated by considering only the changes in the eigenvalues of the correlation operators. To this end, we employ the commutation relations between the CNOT operation and Pauli operators, as described in \cite{Reiter_2011,Wallnofer_2015}. These relations allow us to derive the corresponding transformations of the correlation operators for the specific qubit sets$K_{{\mu}_{T}}, K_{\boldsymbol{\mu}_{T_{N}}}, K_{{\nu}_{T}}, K_{\boldsymbol{\nu}_{T_{N}}}$, defined by Eq.~\eqref{eq:K}. This results in several equations
\begin{equation}
\label{eg:appendixxx}
\begin{gathered}
\begin{aligned}
K_{\nu^{(s)}_{T}}U_{M}\ket{\boldsymbol{\omega}} = \sigma_{x}^{\nu^{(s)}_{T}}\sigma_{z}^{N(\nu^{(s)}_{T})} U_{M}\ket{\boldsymbol{\omega}}= 
\end{aligned}
\\[0.5em]
\begin{aligned}
=U_{M}\sigma_{x}^{\nu^{(s)}_{T}}\sigma_{x}^{\mu^{(t)}_{T}}\sigma_{z}^{\boldsymbol{\nu}^{(t)}_{T_N}}\sigma_{z}^{\boldsymbol{\mu}^{(s)}_{T_N}}\ket{\boldsymbol{\omega}}=
\end{aligned}
\\[0.5em]
\begin{aligned}
    =U_{M}K_{\nu_{T}}K_{\mu_{T}}\ket{\boldsymbol{\omega}}=(-1)^{\nu_{T}\oplus \mu_{T}}U_{M}\ket{\boldsymbol{\omega}}
\end{aligned}
\end{gathered}
\end{equation}
which yields the tensor product of basis states in the graph-state basis. This is also done for each correlation operator, which results in the updated state
\begin{equation}
\label{eg:gffb}
\begin{gathered}
\begin{aligned}
\ket{{\mu}_{T}, \boldsymbol{\mu}_{T_{N}}, \boldsymbol{\mu}_{T_{0}}}
\ket{{\nu}_{T}, \boldsymbol{\nu}_{T_{N}}}
\end{aligned}
\\[0.5em]
\downarrow {}
\\[0.5em]
\begin{aligned}
\ket{{\mu}_{T}, \boldsymbol{\mu}_{T_{N}} \oplus \boldsymbol{\nu}_{B}, \boldsymbol{\mu}_{T_{0}}}
\ket{{\nu}_{T} \oplus {\mu}_{T}, \boldsymbol{\nu}_{T_{N}}}.
\end{aligned}
\end{gathered}
\end{equation}
\section{Fixed Target Fidelity - TCP vs. LEP Resource Scaling}
\label{additional app}

\begin{figure}
    \centering 
    \includegraphics[width=\columnwidth]{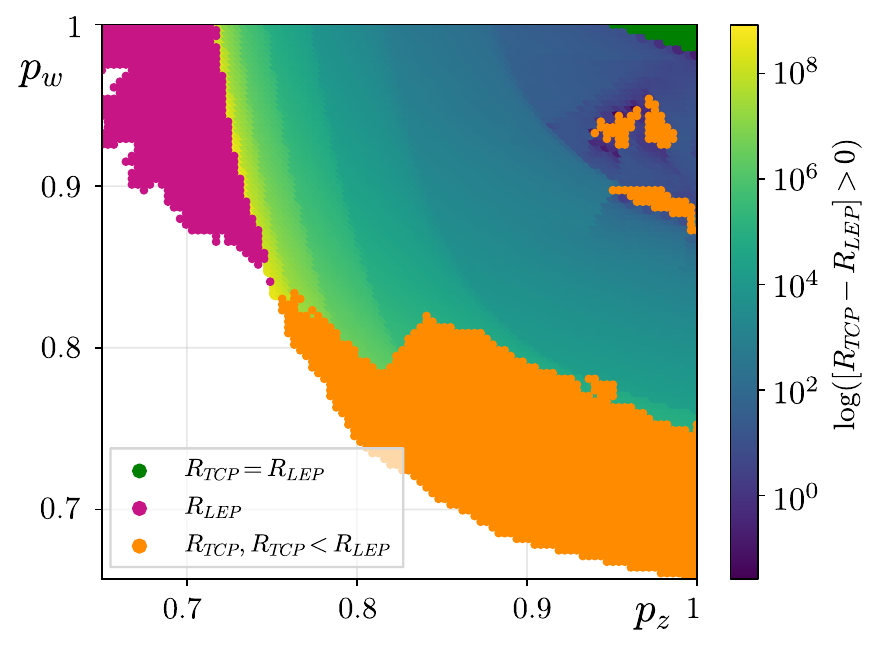}
    \caption{\textbf{TCP vs. LEP Resource Scaling ($F_T$ = 0.90).} Resources difference between LEP and TCP strategy for the fixed target fidelity of $F_T = 0.90$. With varying noise parameters $p_z$ and $p_w$ on qubits 1 and 6. The gate noise is fixed at $p_g = 0.998$. Quantifying the difference in resource requirements between the two protocols as a function of the noise parameters. }
    \label{fig:extra}
\end{figure}

We include an additional plot illustrating the resource difference between LEP and TCP, highlighting TCP's higher resource demands in certain regimes. This serves as a more detailed representation of Fig.~\ref{fig:TF}(b) under the same example conditions (see Sec.~\ref{TF}). The result, shown in Fig.~\ref{fig:extra}, identifies regions where both strategies are applicable; the difference in resource consumption to achieve a fixed target fidelity of $F_T = 0.90$ is highly significant. As illustrated by the quantity
$\text{log}(\left[ R_{TCT} - R_{LEP} > 0 \right])$, the resource advantage of LEP increases markedly with stronger Pauli-$Z$ noise and white noise, demonstrating a substantial reduction in the required resources compared to TCP. In the regime of strong asymmetric noise and low white noise (pink region), LEP becomes the dominant strategy, outperforming TCP without requiring any TCP steps, especially under strong Pauli-$Z$ noise. At very high white-noise levels, TCP dominates without LEP (orange region), albeit at the cost of increased resource consumption starting at $10^4$.

\section{Fixed Target Fidelity - hybrid approach}
\label{ap:TF}

We provide additional figures corresponding to the same scenario discussed in the main text (Sec. \ref{TF}), in which strategies must reach a fixed target fidelity of $ F=0.90$.  Here, we present additional simulation results for a specific hybrid strategy ($LEP-TCP-\alpha$), $S-\alpha$, and TCP. The strategy that consumes the fewest resources is declared the most efficient for the given noise regime.

\begin{figure}
    \subfloat[]{\includegraphics[width=0.45\textwidth]{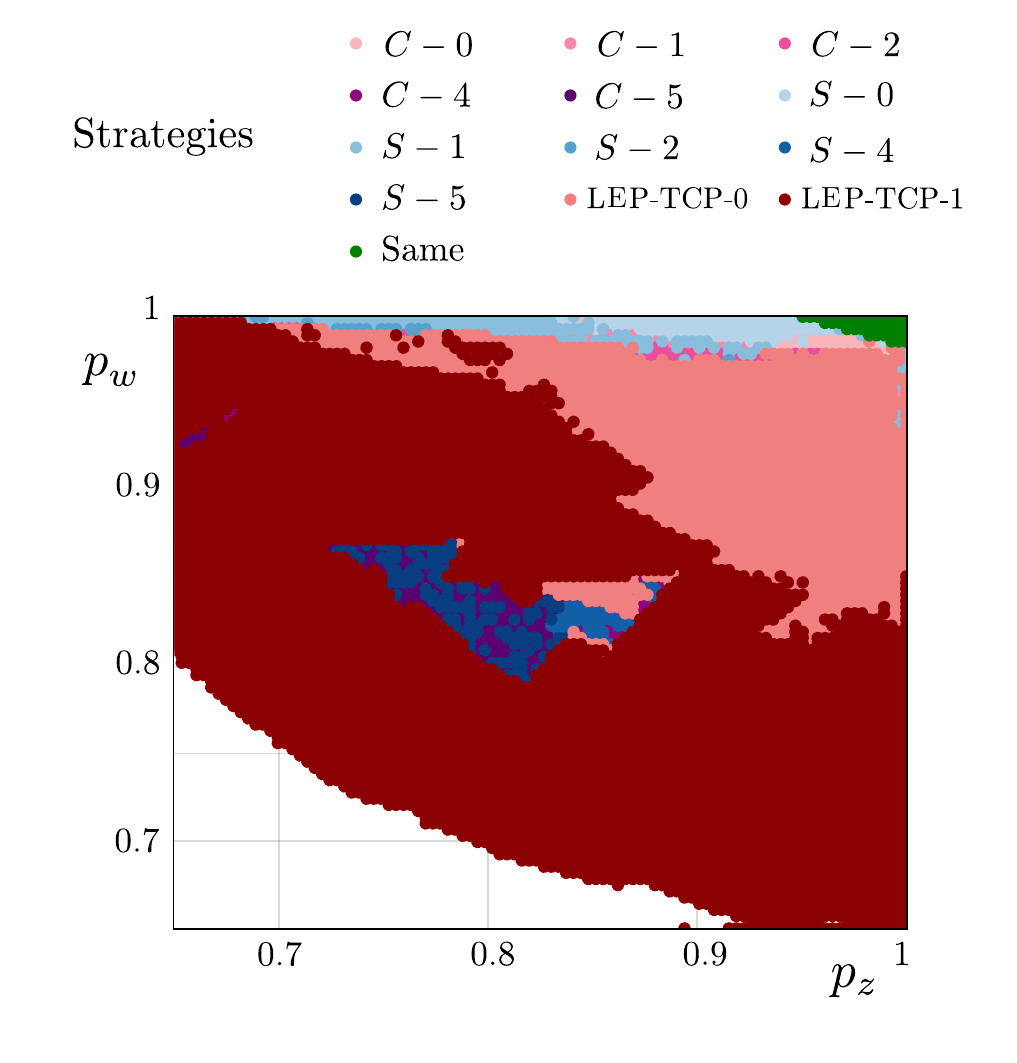}}
    \\[0.5em]
    \subfloat[]{ \includegraphics[width=0.45\textwidth]{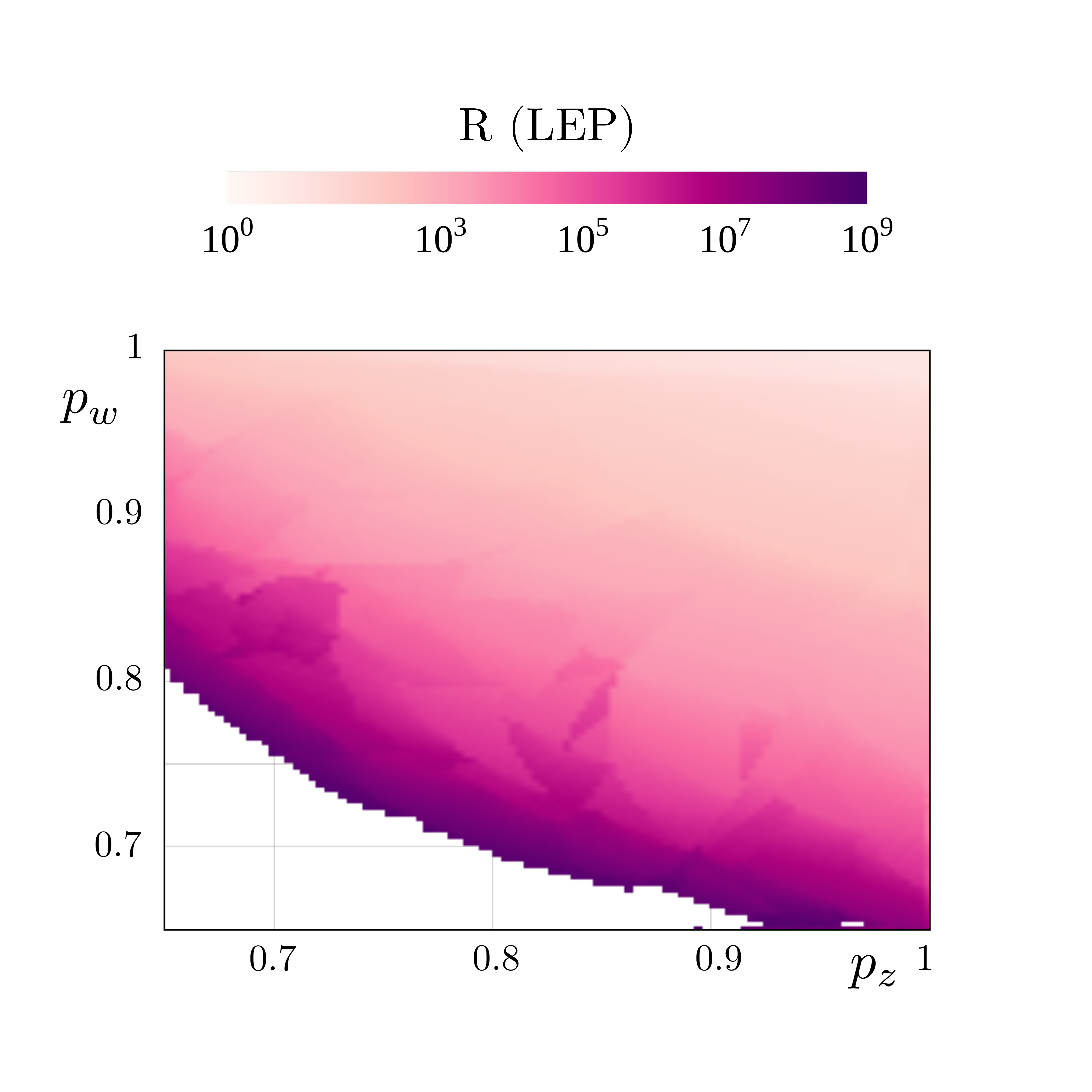}}
    \caption{\textbf{ Hybrid strategy for the fixed target fidelity $F_T = 0.90$.} We consider an 8-qubit linear cluster state subject to white noise with $p_g = 0.998$ and asymmetric $Z$ noise of strength $p_z$ acting on qubits 1 and 6, while the white-noise parameter $p_w$ is varied simultaneously. The simulations compare the strategies S-$\alpha$, C-$\alpha$, TCP, LEP–TCP-$\alpha$, and the same. Sub-figure (a) plots the winning strategies, while sub-figure (b) plots the corresponding resource consumption.}
    \label{fig:Tfap}
\end{figure}

The effects of $S-\alpha$ and TCP were discussed in the main text; we provide further details on the hybrid strategy, which, intuitively, is the most effective approach. Due to the optimized application of $S-\alpha$ for several rounds to mitigate asymmetric noise, thereby transforming it into a more symmetric noise profile, which requires a TCP application. This behavior is clearly evident in Fig. \ref{fig:Tfap}(a), where the hybrid strategy outperforms all others in achieving $F_T$ with the least amount of resources. Consequently, it reaches noisier regimes.

For $F_T=0.90$, certain noise combinations favor the application of $S-\alpha$ or $C-\alpha$ with several purification rounds over the hybrid approach, because the hybrid approach for the first purification round requires a very large amount of resources, as shown in Fig. \ref{fig:2d4cs} for the 2D cluster state example. However, for specific \textit{sweet spots} that require only a small push, it prefers other LEP strategies, such as $S-\alpha$ and $C-\alpha$, which consume less resource. These spots are very adaptive to the fixed target fidelity value. Ultimately, the TCP strategy does not appear optimal.
\begin{figure}
    \subfloat[]{\includegraphics[width=0.45\textwidth]{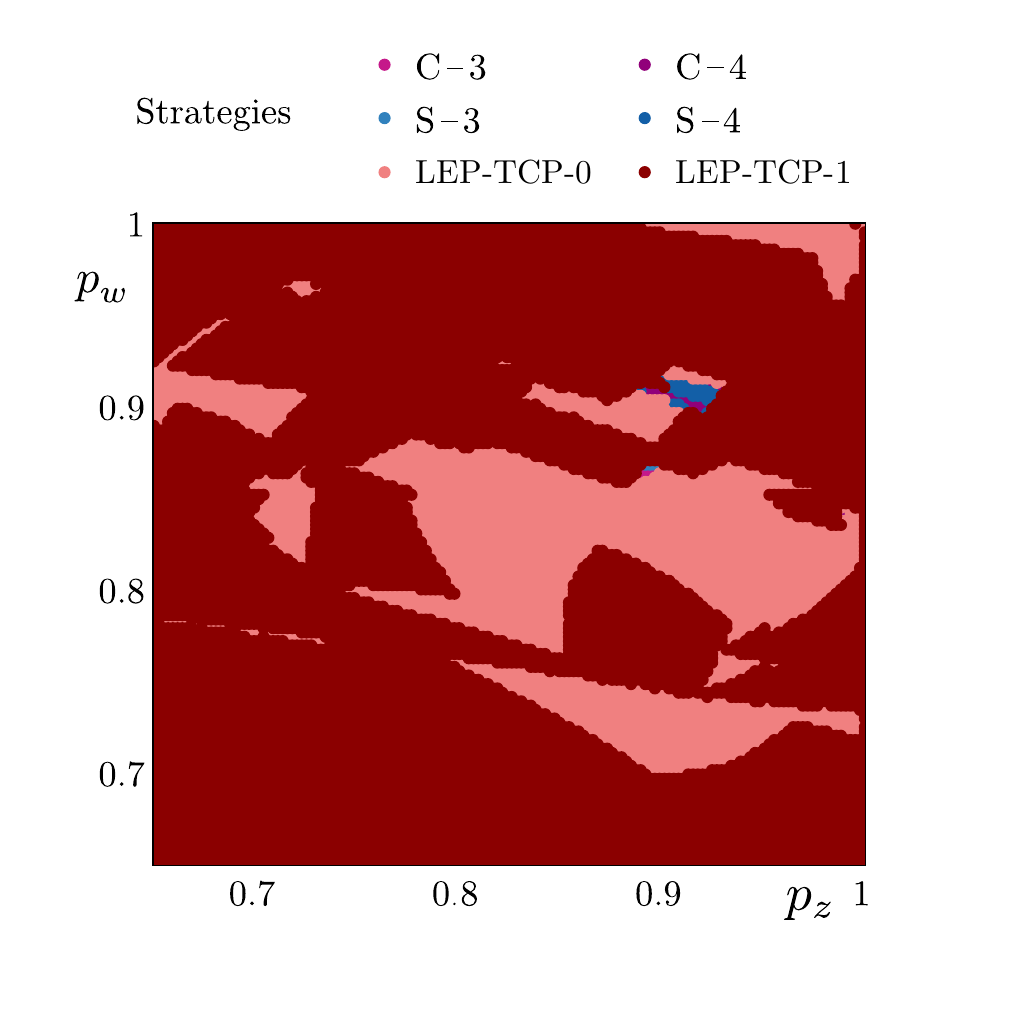}}
    \hfill
    \subfloat[]{ \includegraphics[width=0.45\textwidth]{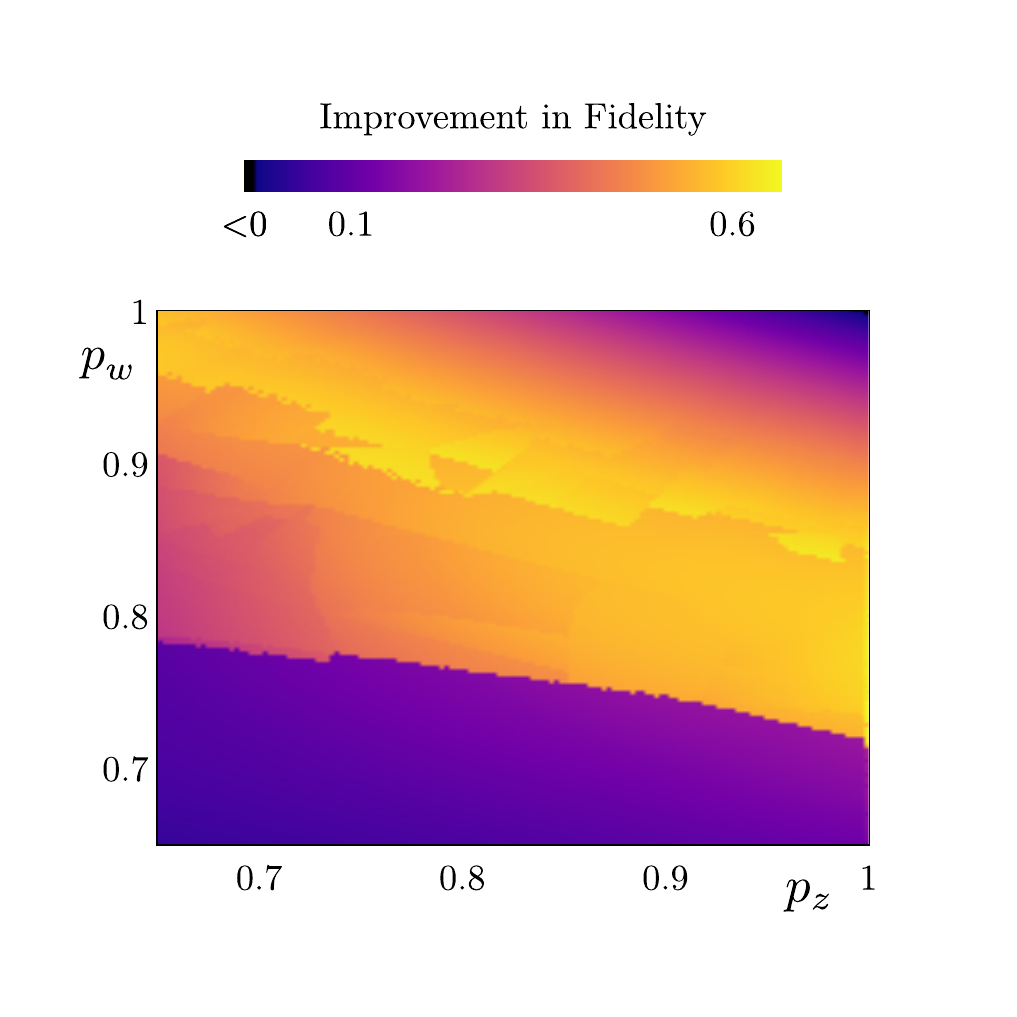}}
    \caption{\textbf{ Hybrid strategy for the Fixed Total Resources $TR = 1000$.} For the case of an 8-qubit linear cluster state subjected to white noise with $p_g = 0.998$, where we consider an asymmetric noise scenario with strength $p_z$ acting on qubits 1 and 6, while the white-noise parameter $p_w$ is varied simultaneously. The simulation uses the following strategies: S-$\alpha$, C-$\alpha$, TCP, and LEP-TCP-$\alpha$. Sub-figure (a) shows the winning strategies for the given noise case, while sub-figure (b) plots the corresponding fidelity improvement of the winning strategy.}
    \label{fig:TRap}
\end{figure}

We also plotted the corresponding amount of resources per strategy, which is illustrated in Fig.~\ref{fig:Tfap}(b). On one hand, the overall resource consumption is equal to the maximum resource consumption done by TCP in Fig. \ref{fig:TF}(b) since we limited for out simulations the resources to $10^9$, on the other hand the hybrid approach reaches further in regions, keeping still the resources lower in those regimes where TCP is maximized by the resources or unreachable.

The hybrid strategy extends the resource threshold into noisier regimes where neither TCP nor LEP alone can efficiently reach the fixed target fidelity.

\section{Fixed Total Resources - hybrid approach}
\label{ap:tr}

In this section, we present additional simulations of the hybrid approach ($LEP-TCP-\alpha$) using established LEPs (S-$\alpha$ and C-$\alpha$) and TCP strategies. 

We investigate the same scenario described in the main text, Sec.~\ref{sec:TR}, with the same restriction of Fixed Total Resource $TR=1000$ to define the approach that yields the highest fidelity under this constraint. 

The Fig.~\ref{fig:TRap}(b) provides the overview of the strategy that yields the highest fidelity at a given noise regime. The hybrid strategy is the most efficient at mitigating noise; thus, it reaches an almost perfect state faster, since it mitigates asymmetric noise with the $S-\alpha$ approach, followed by the TCP step that targets the symmetric noise. Unfortunately, one must account for substantial resource consumption in the first purification step, as illustrated in Fig. \ref{fig:2d4cs}. Since the resource restriction $TR$ is sufficiently large, the hybrid approach eventually dominates overall.

To further assess the usability of the purification protocol, additional information is required. Figure~\ref{fig:TRap}(b), therefore, shows the fidelity difference between the initial state and the state obtained after purification, evaluated either when the $TR$ limit is reached or when the pure state is achieved. The results exhibit a clearer, more pronounced trend, enabling a more straightforward identification of the purification regime and its effectiveness. In particular, this regime indicates either that the purification limit has been reached or that the initial graph state lacks sufficient entanglement for successful purification.

\end{document}